\documentclass[reprint, amsmath, amssymb, aps,
superscriptaddress
]{revtex4-2}

\usepackage{graphicx} 
\usepackage{dcolumn} 
\usepackage{xcolor}
\usepackage{bm}

\usepackage{amsmath}
\usepackage{etoolbox}

\makeatletter
\patchcmd{\endalign}{\restorealignstate@}{\global\let\df@label\@empty\restorealignstate@}{}{}
\makeatother

\usepackage[hidelinks]{hyperref} 

\usepackage{comment}

\newcommand*{\dif}{\mathop{}\!\mathrm{d}}
\def\at{\mathrm{at}}
\def\ss{\mathrm{ss}}
\def\eq{\mathrm{eq}}

\begin{document}
\preprint{APS/123-QED}

\title{Variational quantitative phase-field modeling of non-isothermal sintering process}

\author{Timileyin David Oyedeji}
\email{timileyin.oyedeji@tu-darmstadt.de}
\author{Yangyiwei Yang}
\email{yangyiwei.yang@mfm.tu-darmstadt.de}
\affiliation{
 Mechanics of Functional Materials Division, Institute of Materials Science, Technische Universität Darmstadt, 64287 Darmstadt, Germany}

\author{Herbert Egger}
\affiliation{
 Johann Radon Institute for Computational and Applied Mathematics and Institute for Computational Mathematics, Johannes-Kepler University Linz, 4040 Linz, Austria}

\author{Bai-Xiang Xu}
 \email{xu@mfm.tu-darmstadt.de}
\affiliation{
 Mechanics of Functional Materials Division, Institute of Materials Science, Technische Universität Darmstadt, 64287 Darmstadt, Germany}

\date{\today}

\begin{abstract}
Phase-field modeling has become a powerful tool in describing the complex pore-structure evolution and the intricate multi-physics in non-isothermal sintering processes. However, the quantitative validity of conventional variational phase-field models involving diffusive processes is a challenge. Artificial interface effects, like the trapping effects, may originate at the interface when the kinetic properties of two opposing phases are different.
On the other hand, models with prescribed antitrapping terms do not necessarily guarantee the thermodynamics variational nature of the model.
{This issue has been solved for liquid-solid interfaces via the development of the variational quantitative solidification phase-field model. However, there is no related work addressing the interfaces in non-isothermal sintering, where the free surfaces between the solid phase and surrounding pore regions exhibit strong asymmetry of mass and thermal properties.} Also, additional challenges arise due to the conserved order parameter describing the free surfaces. In this work, we present a variational and quantitative phase-field model for non-isothermal sintering processes. The model is derived via an extended non-diagonal phase-field model. The model evolution equations have naturally cross-coupling terms between the conserved kinetics (i.e., mass and thermal transfer) and the non-conserved one (grain growth). These terms are shown via asymptotic analysis to be instrumental in ensuring the elimination of interface artefacts, while also examined to not modify the thermodynamic equilibrium condition (characterized by dihedral angle). Moreover, we demonstrate that the trapping effects and existence of surface diffusion in conservation laws are direction-dependent. An anisotropic interpolation scheme of the kinetic mobilities which differentiates the normal and the tangential directions along the interface is discussed. Numerically, we demonstrate the importance of the cross-couplings and the anisotropic interpolation via presenting thermal-microstructural evolutions.

\end{abstract}

\maketitle

\section{\label{sec:intro}Introduction}
Sintering is a typical densification technique in thermal processing of bulk materials from packed powders \cite{german2014sintering, kang2004, Niu2010}. In the present day, many new techniques based on sintering have been proposed and broadly applied in the industry,  where the thermal bonding effect is introduced by treatments other than direct heating, such as laser scan, electrical current and electromagnetic field \cite{min2017,Niu2010,Munir2006,Gu2012g}. These techniques are collectively termed as ``unconventional'' sintering \cite{pinto2021overview, yangscripta2020}. 
Due to the distinct heating mechanisms among unconventional sintering techniques, effects of non-isothermal factors on the properties of products, like heating/cooling rate and temperature inhomogeneity gain increasing attentions alongside the conventional ones such as chemical composition as well as size of powders, atmosphere, and pressure. 

Therefore, it is essential to identify and understand the physical effects and interactions of these factors in the sense of bridging the process parameters, microstructure, and properties of the materials to further tailor the performance for applications of interest. Two major types of interfaces are essential for sintering process, namely the free surface between pore and substance and the grain boundary between adjacent crystal grains. There are analytical models for describing the pores/grains evolution, the two-particle coalescence model~\cite{frenkel1945viscous, kuczynski1949self}, dodeca-/tetrakaidecahedron grain model~\cite{smith1948grains}, and various models treating the pores/grains through assumed geometries, like spheres or cylinders~\cite{coble1961sintering, mackenzie1949phenomenological}. Nevertheless, complex grain/pore geometry and entangled multiple physics during sintering goes beyond the capacity of these models. 

For such purpose, phase-field modeling and simulation is promising. In the conventional variational phase-field theory, order parameters (OP) are applied to represent the spatio-temporal distribution of microstructure, i.e. pores and grain orientations in the case of sintering. The thermodynamic potential of the microstructure can then be formulated by an energy functional w.r.t. the OPs, including the interface contribution through the corresponding gradient terms of OPs. From non-equilibrium thermodynamics, the evolution equations of the OPs can be derived on the basis of the variational theory. It circumvents the  necessity of interface tracking. There are variational phase-field sintering models considering an isothermal scenario. For instance, Kazaryan et al.~\cite{Kazaryan1999} and Wang~\cite{wang2006} proposed a line of phase-field model, which was used later for studying two-particle necking and coalescence~\cite{Kumar2010c, ahmed2013, Deng2012, yang2018, Biswas2018c, Biswas2018, Chockalingam2016b} and densification of porous microstructure~\cite{ahmed2013, Ahmed2014c}, and in simulating the overall microstructure evolution of the particle aggregation~\cite{Mukherjee2011} or particle stack~\cite{wang2006, Termuhlen2021}. Rigid-body motions were also incorporated within the model~\cite{wang2006, Termuhlen2021,Biswas2018c,Biswas2018}. Furthermore, a phase-field sintering model adopting the grand potential concept was also developed~\cite{Hotzer2019a, Greenquist2020}. 
To simulate sintering process under highly heterogeneous thermal environment, the phase-field sintering model coupled with transient heat and/or chemical diffusion simulations are needed. The phase-field approach allows such consideration through additive inclusion of the energy contributions by the related physical fields, such as temperature or chemical concentration. In our previous work~\cite{Yang2019}, a variational non-isothermal phase-field sintering model was proposed, which was applied for simulations of the selective laser sintering on a single-layer and multi-layer \cite{zhou2021GAMM} powder beds, and for sintering under prescribed high temperature gradient~\cite{yangscripta2020}. 

On the other hand, one theoretical issue of the conventional variational phase-field models involving thermal/chemical diffusive process is the quantitative validity. Artificial interface effects may originate from violation of conservation laws and discontinuity of the chemical/thermal potentials at the interface (trapping effects)~\cite{Almgren1999, mcfadden2000}. These interface effects scale with the interface width. Theoretically, via asymptotic analysis, phase-field models should be reduced to their associated free-boundary problems in order to ensure their quantitative validity. Based on thin-interface limit analysis, Karma and Rappel~\cite{karma1996, karma1998} first published a quantitative phase-field model for the solidification of pure materials with equal diffusivities in the solid and liquid phases. Moreover, by introducing an antitrapping term in the diffusion flux equation {in order to eliminate the trapping effect, Karma}~\cite{karma2001} presented a quantitative model for the case of isothermal solidification of alloys with negligible diffusivity in the solid phase. Furthermore, for the case with arbitrarily different diffusivities in opposing phases, corresponding antitrapping terms have been also proposed for isothermal \cite{ohno2009} and non-isothermal consideration \cite{Ohno2012}. Thereby a new parameter relating the interface velocity and diffusion flux was further introduced to ensure full elimination of all interface artifacts. 

It should be noted that modifying a variationally derived evolution equation by prescribed antitrapping terms do not necessarily guarantee the variational nature of the model, which is, however, important for thermodynamics soundness. Therefore there have been efforts to develop variational formulations of quantitative phase-field models. Using phenomenological linear relations, variational formulation of quantitative phase-field models have been developed by considering kinetic cross-coupling between the conserved diffusion fields and the nonconserved OPs (non-diagonal model) \cite{brener2012, Boussinot2013, Fang2013, Boussinot2014}. Time evolution equations of the models then exhibit cross-coupling kinetic terms that are formulated in a similar fashion due to Onsager's symmetry. Furthermore, the parameters of these coupling terms are explicitly formulated in terms of the models parameters by considering relations between the models and their sharp-interface counterparts. The cross-coupling term in the diffusion equations which can be likened to the antitrapping term alongside the coupling term in the phase-field evolution equations have been noted to enable full elimination of artificial interface effects \cite{Boussinot2014, Boussinot2017}. The non-diagonal model has been employed to investigate quantitative phase-field simulations of dendritic growth \cite{wang2020} and to examine quantitative simulations of eutectic and eutectoid transformations \cite{wang2021} in which the necessity of the cross-coupling terms were substantiated in both instances.

By separately considering the thermodynamic quantities of two opposing phases and then treating the interface as a mixture of the phases (two-phase variational approach), Ohno et al. ~\cite{ohno2016, ohno2017} presented quantitative variational phase-field models for binary alloy solidification with two-sided diffusion. In the two-phase formulation the {diffusion fields mixture laws} are ensured at the interface as constraints implemented by the Lagrange multiplier approach, and the flux fields of each single-phase fields are formulated variationally. Emergence of Lagrange multipliers in thermodynamic potential formulation gives rise to cross-coupling terms in the model time evolution equations which serve to eliminate the artificial interface effects. Additionally, the necessity of an anisotropic interpolation of the diffusivity (different interpolations for the normal and the tangential directions across the diffuse interface) is demonstrated in eliminating the anomalous interface effects. Though the two-phase variational approach is promising for the study of quantitative validity, the variational nature of the model is only implicitly implemented through variationally formulated single-phase fluxes. The variational behavior of the final model after inserting the Lagrange multiplier still needs to be examined. {Moreover, due to the assumptions of negligible temperature jump or chemical potential jump across the diffuse interface, the models in Refs.~\cite{ohno2016, ohno2017} are applicable mostly for slow solidification processes.}

Based on literature review, there is currently no variational quantitative phase-field model for non-isothermal sintering. In comparison to the non-isothermal solidification models with non-conserved OPs, additional challenge can be expected due to the conserved OPs involved here. In this work, we derive variational formulation of a quantitative phase-field model for non-isothermal sintering processes where the free surfaces between the solid phase and surrounding atmosphere/pore regions have strong asymmetry of both mass and thermal properties. The model is derived via an extension of the non-diagonal phase-field model. Different from the conventional variational non-isothermal sintering phase-field models, the derived model contains cross-coupling terms in the diffusion and phase-field evolution equations, which are essential to ensure the quantitative validity of the model. Furthermore, we demonstrate that the existence of the trapping effects and presence of surface diffusion in conservation laws are direction-dependent. It hence highlights the need of an anisotropic interpolation of the diffusivity tensor.

The paper is structured as follows. The formulations of the quantitative model {(denoted as ``quantitative model'' hereinafter)} are derived in Section \ref{sec:model} where the entropy functional and time evolution equations are explicitly given. Sharp-interface description across solid free surfaces is briefly explained in Section \ref{sec:sharpinterface}. Afterwards, a linkage between model parameters and sharp-interface equations using a reduction procedure is demonstrated in Section \ref{sec:thininterfacelimit}. Section \ref{sec:results} shows the verification and importance of quantitative model followed by comparing with the non-isothermal sintering model proposed in our former work {(denoted as ``existing model'' hereinafter)}. Conclusions are presented in Section \ref{sec:conclusions}.

\section{\label{sec:model}Model Formulation}
Underlying physical processes involved in non-isothermal sintering can be classified as but not limited to: (a) the mass/heat transport, including diffusion through sorts of paths (volume, surface, and grain boundaries) and mass flows (viscous or fluid flow); (b) the structural relaxation, including the rigid-body motions of powders and interface (mostly the grain boundaries) migration. All underlying interactive processes collectively lead to two significant phenomena: one is the densification (eliminating the pores), in which the total surface energy should be reduced; the other is the grain coarsening, in which the total grain-boundary energy should decrease~\cite{german2014sintering, kang2004, olevsky1998theory}. 
In the following, we then present a framework for deriving non-isothermal variational quantitative phase-field sintering model, with its quantitative validity engendered by asymptotic analysis followed. 

\subsection{Entropy and free energy functionals}
In this model, a conserved OP $\rho$ denoting the solid density fraction is used to indicate the solid region ($\rho = 1$) and the atmosphere/pores region ($\rho = 0$) while a series of non-conserved OPs $\{\eta_i\}$ are used to represent the different grain orientations of the solid grains. 
Considering a non-isothermal scenario, the entropy functional $S$ for a subdomain $\Omega$ within the sintering system is defined as
\begin{equation}
\begin{split}
S(e,\rho,\left\{\eta_i\right\}) = & \int_\Omega \bigg[\bigg. s(e,\rho,\left\{\eta_i\right\}) - \frac{\kappa_\rho}{2}\vert \nabla\rho \vert^2 \nonumber \\  & - \frac{\kappa_\eta}{2}\sum_i\vert\nabla\eta_i\vert^2  \label{a1}
\bigg.\bigg] \text{d}\Omega,
\end{split}
\end{equation}
with
\begin{equation}
s = \frac{1+h(\rho)}{2} s_{\ss}(e_\ss) + \frac{1-h(\rho)}{2} s_\at(e_\at)+ s_\text{cf}(\rho,\left\{\eta_i\right\}), \label{a2}
\end{equation}
where $s$ is the local entropy density, $e$ is the internal energy density while $\kappa_\rho$ and $\kappa_\eta$ are the gradient energy coefficients associated with $\rho$ and $\{\eta_i\}$ respectively. $s_{\ss}$ is the solid phase bulk entropy density  and is dependent on the internal energy density of the solid $e_\ss$. The bulk entropy density of the atmosphere $s_{\at}$ is dependent on the internal energy density of the atmosphere $e_\at$. $h(\rho) = 2\rho - 1$ is an interpolation function. The configurational entropy $s_\text{cf}$ is related to the spatial distribution of entropy density proportional to $\rho$ and $\{\eta_i\}$. It is formulated in a form of Landau-type polynomial similar to the one given by Ref.~\cite{wang2006} as
\begin{eqnarray}
s_\text{cf}(\rho,\{\eta_i\}) =&& \underline{C}_\text{cf}\left[\rho^2(1-\rho)^2\right] \: + \underline{D}_\text{cf}\bigg[\bigg.\rho^2 \:+ 6(1-\rho)\sum_i \eta^2_i \: \nonumber \\
&& - 4(2-\rho)\sum_i \eta^3_i \: + 3 \left(\sum_i \eta^2_i \right)^2 \bigg.\bigg] \label{an4},
\end{eqnarray}
where $\underline{C}_\text{cf}$ and $\underline{D}_\text{cf}$ are constants. The multi-well potential in Eq.~(\ref{an4}) can be seen to exhibit minimal at various regions such as: atmosphere ($\rho=0, \: \{\eta_1 = 0, \cdots, \eta_n = 0$\}), and solid grains at different orientations ($\rho= 1, \:\{\eta_1 = 1, \cdots, \eta_n = 0 $\}), $\cdots$ , ($\rho= 1, \:\{\eta_1 = 0, \cdots, \eta_n = 1$\}). One advantage of this potential form is that its constant parameters can be directly linked to material properties \cite{ahmed2013}.

Assuming $e$ can be expressed as
\begin{equation}
e = \frac{1+h(\rho)}{2}e_{\ss} + \frac{1-h(\rho)}{2} e_\at + e_\text{pt}(\rho,\left\{\eta_i\right\}), \label{a3}
\end{equation}
where $e_\text{pt}$ accounts for the spatial distribution of the internal energy proportional to $\rho$ and $\{\eta_i\}$ and is also formulated similar to $s_\text{cf}$ as
\begin{eqnarray}
e_\text{pt}(\rho,\left\{\eta_i\right\}) =&& \underline{C}_\text{pt}\left[\rho^2(1-\rho)^2\right] \: + \underline{D}_\text{pt}\bigg[\bigg.\rho^2 \:+ 6(1-\rho)\sum_i \eta^2_i \: \nonumber \\
&& - 4(2-\rho)\sum_i \eta^3_i \: + 3 \left(\sum_i \eta^2_i \right)^2 \bigg.\bigg],
\label{an2}
\end{eqnarray}
where $\underline{C}_\text{pt}$ and $\underline{D}_\text{pt}$ are constants.

Following the Legendre transformation, we can obtain the free energy functional $F$ as 
\begin{eqnarray}
F (T,\rho,\left\{\eta_i\right\}) =&& \int_\Omega \bigg[\bigg. f(T,\rho,\left\{\eta_i\right\}) + \frac{T\kappa_\rho}{2}\vert \nabla\rho \vert^2 \nonumber \\&& + \frac{T\kappa_\eta}{2}\sum_i\vert\nabla\eta_i\vert^2 
\bigg.\bigg] \text{d}\Omega, 
\label{an1}
\end{eqnarray}
with
\begin{equation}
f(T,\rho,\left\{\eta_i\right\}) = \frac{1+h(\rho)}{2}f_\ss(T) + \frac{1-h(\rho)}{2}f_\at(T) + e_\text{pt} - Ts_\text{cf}, \label{an3}
\end{equation}
where $f_{\ss}$ and $f_{\at}$ are the free energy densities of the solid phase and the atmosphere, respectively. $T$ is the temperature. Substituting Eqs.~(\ref{an4}) and (\ref{an2}) into Eq.~(\ref{an3}), we obtain
\begin{eqnarray}
f(T,\rho,\left\{\eta_i\right\}) =&&  \frac{1+h(\rho)}{2}f_\ss(T) + \frac{1-h(\rho)}{2}f_\at(T) \nonumber \\ &&+ \underline{C}\left[\rho^2(1-\rho)^2\right]  + \underline{D}\bigg[\bigg.\rho^2 + 6(1-\rho)\sum_i \eta^2_i \: \nonumber \\ && - 4(2-\rho)\sum_i \eta^3_i + 3 \left(\sum_i \eta^2_i \right)^2 \bigg.\bigg] \label{nan1},
\end{eqnarray}
with
\begin{eqnarray*}
\underline{C}&=& \underline{C}_\text{pt} - T\underline{C}_\text{cf}, \\
\underline{D} &=& \underline{D}_\text{pt} - T\underline{D}_\text{cf}.
\end{eqnarray*}

\subsection{Kinetic equations}
Considering that $\rho$ and $e$ are conserved OPs, they satisfy mass and energy conservation laws respectively:
\begin{equation}
\dot{\rho} = - \nabla \cdot \mathbf{J}_{\rho}, \label{a4}
\end{equation}
\begin{equation}
\dot{e} = - \nabla \cdot \mathbf{J}_{e}, \label{a5}
\end{equation}
where $\mathbf{J}_{\rho}$ is the mass diffusion flux and $\mathbf{J}_{e}$ is the energy flux.

Following our previous work \cite{yangscripta2020}, the non-negative entropy production $\sigma$ in the subdomain can be formulated as
\begin{eqnarray}
\sigma = \int_\Omega \bigg[\mathbf{J}_{\rho}\cdot\nabla \frac{\delta S}{\delta \rho} + \mathbf{J}_{e}\cdot\nabla \frac{\delta S}{\delta e} + \sum_i \dot{\eta_i}\frac{\delta S}{\delta \eta_i} \bigg]\text{d}\Omega, \label{a6}
\end{eqnarray}
with
\begin{equation*}
\frac{\delta S}{\delta \rho}= -\frac{1}{T}\frac{\delta F}{\delta \rho}, \: \frac{\delta S}{\delta \eta_i}= -\frac{1}{T}\frac{\delta F}{\delta \eta_i}, \: \frac{\delta S}{\delta e}= \frac{1}{T},
\end{equation*}
where $\nabla ({\delta S}/{\delta \rho})$ is the driving force associated with $\mathbf{J}_{\rho}$, $\nabla ({\delta S}/{\delta e})$ is the driving force associated with $\mathbf{J}_{e}$ and ${\delta S}/{\delta \eta_i}$ is the driving force associated with $\dot{\eta_i}$. 

In the view of the phenomenological linear laws of non-equilibrium thermodynamics and also ensuring non-negative production of the entropy, we can define the relationships between the fluxes, the non-conserved OPs time evolution equations and their driving forces as
\begin{center}
\begin{equation}
\begin{bmatrix} 
\mathbf{J}_{\rho} \\
\mathbf{J}_{e}\\
\dot{\eta_1}\\
\vdots   \\
\dot{\eta_n}
\end{bmatrix}
= \begin{bmatrix} 
\mathbf{L}_{\rho\rho} & \mathbf{L}_{\rho e} & \mathbf{L}_{\rho \eta_1}& \cdots & \mathbf{L}_{\rho \eta_n} \\
\mathbf{L}_{e\rho} & \mathbf{L}_{e e} & \mathbf{L}_{e \eta_1}& \cdots & \mathbf{L}_{e \eta_n} \\
\mathbf{L}_{\eta_1 \rho} & \mathbf{L}_{\eta_1 e} & \mathbf{L}_{\eta_1 \eta_1} & \cdots & \mathbf{L}_{\eta_1 \eta_n} \\
\vdots & \vdots & \vdots & \ddots & \vdots \\
\mathbf{L}_{\eta_n \rho} & \mathbf{L}_{\eta_n e} & \mathbf{L}_{\eta_n \eta_1} & \cdots & \mathbf{L}_{\eta_n \eta_n}
\end{bmatrix}
\begin{bmatrix} 
 -\nabla(\frac{\mu}{T})\\
\nabla (\frac{1}{T}) \\
\frac{\delta S}{\delta \eta_1} \\
\vdots    \\
\frac{\delta S}{\delta \eta_n}
\end{bmatrix},
 \label{a7}  
\end{equation}
\end{center}
where $\mu = {\delta F}/{\delta \rho}$ is defined as the chemical potential and $n$ represents the total number of grain orientations. $\mathbf{L}_{\rho\rho}$, $\mathbf{L}_{\rho e}$, $\mathbf{L}_{e\rho}$ and $\mathbf{L}_{ee}$ are positively defined rank 2 tensors and for $i = 1, 2, \cdots, n$ , $\mathbf{L}_{\rho \eta_i}$, $\mathbf{L}_{e \eta_i}$, $\mathbf{L}_{\eta_i \rho}$ and $\mathbf{L}_{\eta_i, e}$ are positively defined rank 1 tensors while $\mathbf{L}_{\eta_i \eta_i}$ is a positively defined rank 0 tensor.

Diagonal terms $\mathbf{L}_{\rho \rho}$ and $\mathbf{L}_{e e}$ are the diffusional mobilities of mass and energy diffusion respectively. Mobility term associated with the grain orientations, $\mathbf{L}_{\eta_i \eta_i}$ is simply a scalar function and is thereafter taken as $L_\eta$ where we assume isotropic condition taking it to be the same regardless of $i$. The non-diagonal terms in the Onsager matrix in Eq.~($\ref{a7}$) represent cross-couplings between the various OPs. Based on the Onsager reciprocal relations, we have $\mathbf{L}_{\rho e} = \mathbf{L}_{e\rho}$, $\mathbf{L}_{\rho \eta_i} = \mathbf{L}_{\eta_i \rho}$ and $ \mathbf{L}_{e \eta_i} = \mathbf{L}_{\eta_i e}$. Note that the cross-coupling between the different grain orientations is not considered, resulting in similar $\dot{\eta}_i$ formulation for all $i$. Hence we consider only one $\dot{\eta}_i$ whose formulation is representative for all $i$. The quantities $\mathbf{L}_{\rho e}$ and $\mathbf{L}_{e\rho}$ are associated with the mass flux due to temperature gradient (thermophoresis effect) and with the energy flux due to chemical potential gradient (Dufour effect), respectively. Examination of these effects has been done in our previous work \cite{yangscripta2020} and is not the main priority of this work. Therefore, the terms associated with $\mathbf{L}_{\rho e}$ and $\mathbf{L}_{e\rho}$ in the fluxes formulations are dropped. The time evolution equations can then be written as
\begin{subequations}
\label{an17}
\begin{equation}
\dot{\rho} = \nabla \cdot \bigg[\mathbf{L}_{\rho\rho}\cdot \nabla \bigg(\frac{\mu}{T}\bigg) \bigg] +  \nabla \cdot \bigg[\frac{1}{T}\sum_i \mathbf{L}_{\rho \eta_i} \frac{\delta F}{\delta \eta_i} \bigg] , \label{an17a}
\end{equation}
\begin{equation}
\dot{e} = \nabla \cdot \bigg[\mathbf{L}_{e e} \cdot  \frac{\nabla T}{T^2} \bigg] + \nabla \cdot \bigg[\frac{1}{T}\sum_i \mathbf{L}_{e \eta_i}\frac{\delta F}{\delta \eta_i} \bigg], \label{an17b}     
\end{equation}
\begin{equation}
\dot{\eta}_i = - \mathbf{L}_{\eta_i \rho}\cdot\nabla \bigg(\frac{\mu}{T}\bigg) - \mathbf{L}_{\eta_i, e}\cdot\frac{\nabla T}{T^2} - L_\eta\frac{1}{T}\frac{\delta F}{\delta \eta_i}. \label{an17c}
\end{equation}
\end{subequations}

Formulations expressed in Eqs.~(\ref{a7}) and (\ref{an17}) present the fluxes and time evolution equations of the associated OPs in terms of the driving forces. However, for consistency with previous non-diagonal models \cite{Boussinot2013, Boussinot2014} as well as ease of relating our model to the sharp-interface counterpart as will be discussed later, we reformulate the phenomenological linear relations employing the linear relations of the driving forces in terms of the fluxes and time evolution equations such that
\begin{subequations}
\label{a9}
\begin{equation}
-\nabla\left(\frac{\mu}{T}\right) = \mathbf{L}_{\rho\rho}^{-1}\cdot\mathbf{J}_{\rho} + \mathbf{L}_{\rho \eta_i}^{-1} \sum_i\dot{\eta_i}, \label{a9:1}
\end{equation}
\begin{equation}
-\frac{\nabla T}{T^2} = \mathbf{L}_{ee}^{-1}\cdot \mathbf{J}_{e} + \mathbf{L}_{e \eta_i}^{-1} \sum_i\dot{\eta_i}, \label{a9:2}
\end{equation}
\begin{equation}
-\frac{1}{T}\frac{\delta F}{\delta \eta_i} = \mathbf{L}_{\eta_i \rho}^{-1}\cdot \mathbf{J}_{\rho} + \mathbf{L}_{\eta_i, e}^{-1}\cdot \mathbf{J}_{e} + L_\eta^{-1}\dot{\eta_i}. \label{a9:3}
\end{equation}
\end{subequations}

Since the variation of mass density and internal energy is found across free surfaces of the solid grains, the cross-coupling terms $\mathbf{L}^{-1}_{\rho \eta_i} = \mathbf{L}^{-1}_{\eta_i \rho}$ and $\mathbf{L}^{-1}_{e \eta_i} = \mathbf{L}^{-1}_{\eta_i e}$ should be defined such that they are only evaluated at the free surfaces. Also, the non-equilibrium effects associated with these cross terms scale with the diffuse interface width $l$. Accordingly, following Refs. \cite{brener2012,Boussinot2013,Boussinot2014}, we propose the following formulations:
\begin{subequations}
\label{a8}
\begin{equation}
\mathbf{L}_{\rho \eta_i}^{-1} = \mathbf{L}_{\eta_i \rho}^{-1}= {M}_{1}(\rho)l\nabla\rho, \label{a8:1}
\end{equation}
\begin{equation}
\mathbf{L}_{e \eta_i}^{-1} = \mathbf{L}_{\eta_i e}^{-1} = {M}_{2}(\rho)l\nabla\rho,
\end{equation}
\end{subequations}
where ${M}_1$ and ${M}_2$ are scalar functions used to parametrize the associated cross-coupling terms. $l\nabla\rho$ is a vector normal to the free surfaces and has a magnitude of 1 at the center of the free surfaces assuming the parameter $\alpha$ used to adjust the definition of $l$ in Ref. \cite{Kim1999} equals 2 \cite{Yang2019}. Substituting Eq.~($\ref{a9}$) into ($\ref{a6}$) and taking into account the aforementioned, we obtain the entropy production in the subdomain as
\begin{eqnarray}
\sigma =&& \int_\Omega \bigg[\mathbf{L}_{\rho\rho}^{-1}\cdot\mathbf{J}_{\rho}\cdot\mathbf{J}_{\rho} + \mathbf{L}_{ee}^{-1}\cdot\mathbf{J}_{e}\cdot\mathbf{J}_{e} + L_\eta^{-1}\bigg(\sum_i\dot{\eta_i}\bigg)^2 \nonumber \\  &&+ 2l\nabla\rho\sum_i\dot{\eta_i}\cdot\left(M_{1}\mathbf{J}_{\rho} + M_{2}\mathbf{J}_{e}\right) \bigg]\text{d}\Omega. \label{a12}
\end{eqnarray}
Furthermore, time evolution equations can be obtained as 

\begin{subequations}
\label{ann4}
\begin{equation}
\dot \rho = \nabla \cdot \bigg[ \mathbf{L}_{\rho\rho}\cdot\bigg(\nabla \left(\frac{\mu}{T}\right)  \: + M_{1}l\nabla\rho\sum_i\dot{\eta_i} \bigg) \bigg], \label{ann4a}
\end{equation}
\begin{eqnarray}
c_\text{r}\dot T + \frac{\partial e}{\partial \rho}\dot \rho + \sum_i\frac{\partial e}{\partial \eta_i}\dot{\eta_i} = && \nabla\cdot\bigg[\mathbf{L}_{ee}\cdot\bigg(\frac{\nabla T}{T^2} \nonumber \\ &&+  M_{2}l\nabla\rho\sum_i\dot{\eta_i}\bigg) \bigg],\label{an5}
\end{eqnarray}
\begin{eqnarray}
\hat L_\eta^{-1}\dot{\eta_i} =&& \: \kappa_\eta\nabla^2\eta_i - \frac{1}{T}\frac{\partial f}{\partial \eta_i} \: + \: l\nabla\rho\cdot\bigg[M_1\mathbf{L}_{\rho\rho}\cdot\nabla \left(\frac{\mu}{T}\right)  \: \nonumber \\  &&+ M_2\mathbf{L}_{ee}\cdot\frac{\nabla T}{T^2}\bigg], \label{an6}
\end{eqnarray}
\end{subequations}
with
\begin{equation}
\hat L_\eta^{-1} = L_\eta^{-1} - [M^2_1 l^2\nabla\rho \cdot\mathbf{L}_{\rho\rho}  + M_2^2 l^2\nabla\rho \cdot \mathbf{L}_{ee}]\cdot \nabla\rho. \label{an7} 
\end{equation}

Hereby $c_\text{r} = \frac{1+h(\rho)}{2}c_\ss + \frac{1-h(\rho)}{2}c_\at$ is the relative specific heat, where $c_\ss={\partial e_{\ss}}/{\partial T}$ and $c_\at={\partial e_{\at}}/{\partial T}$ are the volumetric specific heat of solid and atmosphere, respectively.

Comparing the heat transfer equation (Eq.~($\ref{an5}$)) to that of conventional quantitative phase-field model \cite{Ohno2012}, the second term on the right-hand-side (RHS) can be likened to the thermal antitrapping current related to the elimination of thermal trapping (associated with temperature jump) at the free surfaces. Similarly, the second term on the RHS of Eq.~($\ref{ann4a}$) represents some form of antitrapping current valued only at the free surfaces. Similar to solutal antitrapping current~\cite{karma2001, echebarria2004} associated with solute trapping due to jump of chemical potential, this term is termed as the mass antitrapping current in this work. The last two terms on the RHS of the grain orientation time evolution equations (Eq.~($\ref{an6}$)) represent cross-coupling terms associated with mass and energy diffusion across the free surfaces, respectively. These terms alongside the antitrapping terms are absent in time evolution equations of conventional non-isothermal phase-field sintering models but are very vital in the elimination of artificial interface effects such as the trapping effects at the free surfaces of the solid phase.

Moreover, considering no variation of solid density and thermal properties across the grain boundaries, Eq.~($\ref{an6}$) has no cross-coupling terms and simply takes a form of Allen-Cahn equation at the grain boundaries. Consequently, we limit our subsequent analysis and derivations to the free surfaces where the cross-coupling terms are significant.

\section{\label{sec:sharpinterface}Sharp-interface description across free surfaces}
Considering a simple nonisothermal system consisting a sharp free surface between a solid grain and the atmosphere, the following set of sharp-interface equations can be described in the bulk regions:

\begin{equation}
    \frac{\partial \rho}{\partial t} = \nabla \cdot ({M}_\text{rg} \nabla \mu),  \label{si1}
\end{equation}
\begin{equation}
    c_\text{rg}\frac{\partial T}{\partial t} = \nabla \cdot ({k}_\text{rg} \nabla T), \label{si4}
\end{equation}
where for a bulk region $\text{rg}$ ("$\ss$" for solid and "$\at$" for atmosphere), ${M}_\text{rg}$, $c_\text{rg}$, and ${k}_\text{rg}$ represent the region's effective mass mobility coefficient, volumetric specific heat, and effective thermal conductivity respectively. {$\rho$ here adopts the physical meaning of normalised density of the solid. Eqs.~($\ref{si1}$) and (\ref{si4}) describe mass and heat transfer in the bulk regions.  For the bulk atmosphere region in particular, ${M}_\at$ describes the effective mobility considering mass transfer mechanisms notably evaporation and condensation. Hence, the driving force $\nabla \mu$ for mass transfer in the atmosphere takes into account vapor pressure differences due to local curvature \cite{german2014sintering, kang2004}. Similarly, ${k}_\text{at}$ describes effective thermal conductivity taking into account convection and radiation.}

Furthermore, energy conservation condition at the free surface can be described as
\begin{equation}
\begin{aligned}
     v e_{\ss}+ {k}_\ss \left.\nabla T\right\vert_\ss\cdot \mathbf{n}_\text{sf} = v e_{\at} + {k}_\at \left.\nabla T\right\vert_\at\cdot \mathbf{n}_\text{sf} = J_T , \label{si5}
\end{aligned}
\end{equation}
where $v$ is the velocity of the migrating free surface, and $\left.\nabla T\right\vert_\ss$ and $\left.\nabla T\right\vert_\at$ are the spatial gradients of the temperature at the solid and atmosphere sides of the free surface respectively. $\mathbf{n}_\text{sf}$ is the unit vector normal to the free surface. $J_T$ is the normal heat flux flowing through the free surface. 
{Similarly, explicit formulation of mass conservation at the free surface is given as
\begin{equation}
\begin{split}
     v(\rho_{\ss} - \rho_{\at}) =& - {M}_\ss \left.\nabla \mu\right\vert_\ss\cdot \mathbf{n}_\text{sf} 
     + {M}_\at \left.\nabla \mu\right\vert_\at\cdot \mathbf{n}_\text{sf} \\ &+ {M}_\text{sf}\nabla^{2}_\text{sf}\mu, \label{sinw}
\end{split}
\end{equation}
where $\rho_\ss$ and $\rho_\at$ are the bulk densities in the solid and atmosphere, and $\left.\nabla \mu\right\vert_\ss$ and $\left.\nabla \mu\right\vert_\at$ are the spatial gradients of the chemical potential at the solid and atmosphere sides of the free surface respectively. $M_\text{sf}$ represents surface diffusion mobility. $\nabla_\mathrm{sf}^2$ is surface Laplacian. The last term in Eq.~(\ref{sinw}) describes surface diffusion typical to sharp-interface description of mass transfer in sintering \cite{Deng2012}.}
Moreover, $v$ can be defined as
\begin{equation}
    v = v_s + v_b , \label{si2}
\end{equation}
where $v_s$ and $v_b$ are the velocities contributed by surface diffusion and bulk/volume diffusion respectively
and can be expressed in terms of their corresponding mass fluxes;
\begin{equation}
   v_s = - V_\mathrm{m} \nabla_\mathbf{sf}\cdot \mathbf{J}_\text{sf},  \; v_b = - V_\mathrm{m} \mathbf{J}_b \cdot \mathbf{n}_\text{sf},
\end{equation}
where $V_\mathrm{m}$ is the molar volume and $\nabla_\mathbf{sf}$ is the surface gradient. $\mathbf{J}_\text{sf}$ is the mass flux along the free surface associated with surface gradient of the free surface curvature $\mathrm{k}_\text{sf}$; $\mathbf{J}_\text{sf} \propto \nabla_\mathbf{sf} \mathrm{k}_\text{sf}$. $\mathbf{J}_b$ is mass flux from the solid bulk to the free surface associated with the gradient of the chemical potential in the solid bulk grains $\mu_\ss$; $\mathbf{J}_b \propto \nabla \mu_\ss$,  \cite{Deng2012, Maximenko2004}. 

In addition, the chemical potential and temperature at the free surface obey the following relations:
\begin{subequations}
\begin{equation}
\mu\vert_\ss = \mu\vert_\at , \label{si3}
\end{equation}
\begin{equation}
T\vert_\ss - T\vert_\at = J_T R_s  , \label{si6}
\end{equation}
\end{subequations}
where $\mu\vert_\ss$ and $\mu\vert_\at$ represent chemical potentials at the solid and atmosphere sides of the free surface, respectively. $T\vert_\ss$ and $T\vert_\at$ represent the temperatures at the solid and atmosphere sides of the free surface, respectively. $R_s$ represents Kapitza-type thermal resistance. In this work, we assume negligible $R_s$, thereby Eq.~($\ref{si3}$) and Eq.~($\ref{si6}$) indicate imposed zero chemical potential and temperature jumps at the free surface.

Moreover, we infer that jump in chemical potential $\delta \mu$ across the free surface is conjugated to $v$ and also that the temperature jump $\delta T$ across the free surface is conjugated to $J_T$. The kinetic boundary conditions can then be expressed in the framework of phenomenological linear relations as \cite{Balibar2005, BrenerTemkin2012}
\begin{equation}
\delta \mu = \mathcal{A}v + \mathcal{B}J_T,  \label{a14}
\end{equation}
\begin{equation}
\delta T = \mathcal{B}v + \mathcal{C}J_T,  \label{a15}
\end{equation}
where $\mathcal{A}, \mathcal{B}$ and  $\mathcal{C}$ are kinetic coefficients of the positive-definite Onsager matrix. Entropy production at the free surface $\sigma_{s}$ can be formulated as
\begin{equation}
\sigma_{s} = v \delta \mu  \: + J_T\delta T. \label{a17}
\end{equation}

\noindent Substituting Eqs.~($\ref{a14}$) and ($\ref{a15}$) into ($\ref{a17}$), we obtain
\begin{equation}
\sigma_{s} = \mathcal{A}v^2 + \mathcal{C}J_T^2 + 2\mathcal{B}vJ_T. \label{a18}
\end{equation}

\begin{widetext}

\section{\label{sec:thininterfacelimit}Thin interface limit: Linking model with sharp-interface description}
In this section, still considering a system consisting a free surface between a solid grain and the atmosphere, we establish the relationships between $\mathcal{A}$, $\mathcal{B}$ and $\mathcal{C}$ and the phase-field parameters following the reduction procedure presented in Ref. \cite{Boussinot2013}. Considering a 1D system with the free surface centered at $x=0$ (shown in Fig.~\ref{fig:asympscheme}a), we have $\rho$ and $\eta$ vary from a semi-finite solid region ($-\infty$) to a semi-finite atmosphere region ($+\infty$). For simplicity, notation $(\cdot)'$ is adopted to represent the derivative w.r.t. the spatial coordinate $x$. {It is worth noting that we consider the profile of $\rho$ between two bulk values that are slightly deviated from the ideal ones, i.e., $\rho_\ss$ in the substance and $\rho_\at$ in the pore/atmosphere. The origin and the thermodynamic outcome of these deviated bulk values of $\rho$ are explicitly examined and discussed in the {Appendix}.}

\begin{figure}[!h]
\includegraphics[scale=0.6]{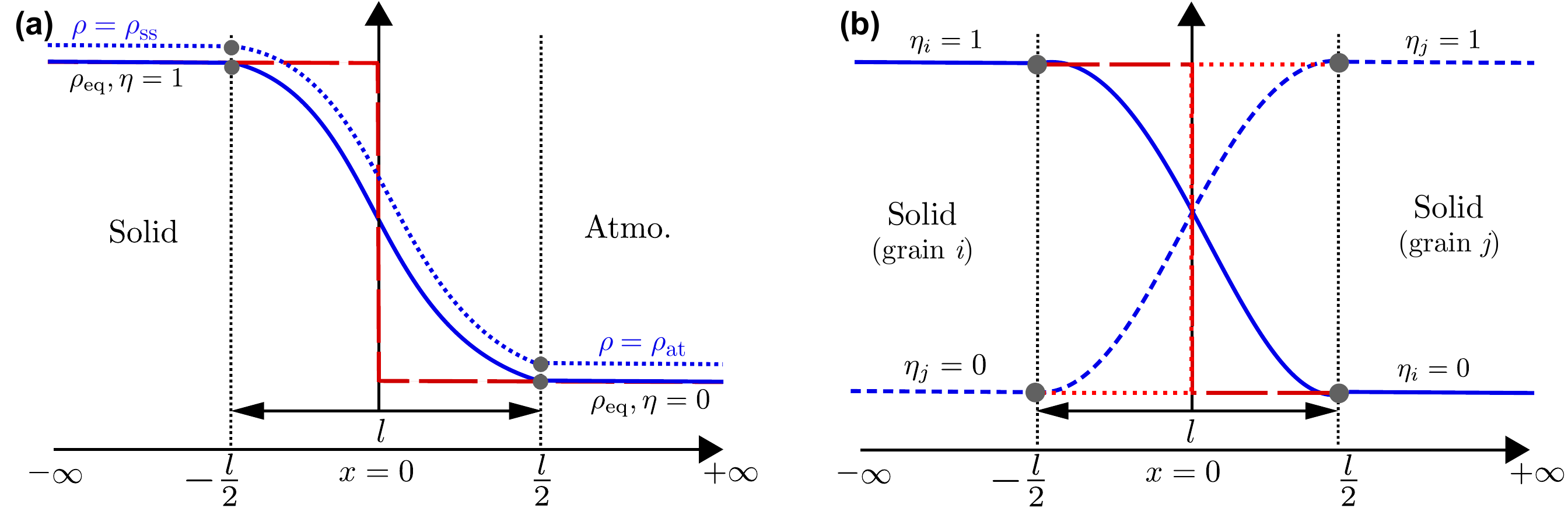}
\caption{\label{fig:asympscheme} (a) Asymptotic schematic across a free surface; blue solid line represents phase-field profile at equilibrium and red large-dashed line represents sharp-interface profile; blue dotted line shows the profile of $\rho$ with slightly deviated values; (b) Asymptotic schematic across a grain boundary; blue solid and dashed lines represent phase-field profiles and red large-dashed and dotted lines represent sharp-interface profiles.} 
\end{figure}

\noindent According to the phase-field method, the entropy production (Eq.~($\ref{a12}$)) for the system considered can be formulated as
\begin{eqnarray}
\sigma =&& \int^{-l/2}_{-\infty} \bigg[\frac{J_\rho^2(x)}{{L}^\ss_{\rho\rho}} + \frac{J_e^2(x)}{{L}^\ss_{ee}} \bigg] \dif x + \int^{\infty}_{l/2} \bigg[\frac{J_\rho^2(x)}{{L}^\at_{\rho\rho}} + \frac{J_e^2(x)}{{L}^\at_{ee}} \bigg] \dif x  \\
&&+\int^{l/2}_{-l/2} \bigg[\frac{J_\rho^2(x)}{{L}_{\rho\rho}} + \frac{J_e^2(x)}{{L}_{ee}} \nonumber + L_{\eta,\text{sf}}^{-1}\dot{\eta}^2 + 2l\rho^{'}(x)\dot{\eta}\left(M_{1}J_\rho(x) + M_{2}J_e(x)\right)\bigg] \dif x, \label{a19}
\end{eqnarray}
where ${L}^\ss_{\rho\rho} = {L}_{\rho\rho}(\rho=\rho_\ss)$ and ${L}^\at_{\rho\rho} = {L}_{\rho\rho}(\rho=\rho_\at)$ are the effective mass mobilities in the corresponding regions. Also, ${L}^\ss_{ee} = {L}_{ee}(\rho=\rho_\ss)$ and ${L}^\at_{ee} = {L}_{ee}(\rho=\rho_\at)$ are the effective energy mobilities in the corresponding regions. $L_{\eta,\text{sf}}$ is mobility of $\eta$ at the free surface. It can be noted that in the bulk regions $(| x | > l/2 )$, only the fluxes $J_\rho(x)$ and $J_e(x)$ contribute to entropy production as $\dot{\eta}$ and $\rho^{'}(x)$ both go to zero. The entropy production of the system considered can be formulated within the sharp-interface description as
\begin{equation}
 \int^{0}_{-\infty} \bigg[\frac{J_\rho^2(x)}{{L}^\ss_{\rho\rho}} + \frac{J_e^2(x)}{{L}^\ss_{ee}} \bigg] \dif x + \int^{\infty}_{0} \bigg[\frac{J_\rho^2(x)}{{L}^\at_{\rho\rho}} + \frac{J_e^2(x)}{{L}^\at_{ee}} \bigg] \dif x + \sigma_s. \label{a20}  
\end{equation}

\noindent Comparing Eqs.~($\ref{a19}$) and ($\ref{a20}$), we obtain entropy production at the free surface within the phase-field model as
\begin{eqnarray}
\sigma_s =&& \int^{l/2}_{-l/2} \bigg[\frac{J_\rho^2(x)}{{L}_{\rho\rho}} + \frac{J_e^2(x)}{{L}_{ee}} + L_{\eta,\text{sf}}^{-1}\dot{\eta}^2 + 2l\rho^{'}(x)\dot{\eta}\bigg(M_{1}J_\rho(x) + M_{2}J_e(x)\bigg)\bigg] \dif x \nonumber \\ &&-\int^{0}_{-l/2} \bigg[\frac{J_{\ss(\rho)}^2}{{L}^\ss_{\rho\rho}} + \frac{J_{\ss(e)}^2}{{L}^\ss_{ee}} \bigg] \dif x - \int^{l/2}_{0} \bigg[\frac{J_{\at(\rho)}^2}{{L}^\at_{\rho\rho}} + \frac{J_{\at(e)}^2}{{L}^\at_{ee}} \bigg] \dif x, \label{a21}
\end{eqnarray}
where for a region $\text{rg}$ (``$\ss$'' for solid and ``$\at$'' for atmosphere), $J_{\text{rg}(\rho)}$ and $J_{\text{rg}(e)}$ represent the region's bulk mass and energy fluxes respectively.

For the purpose of making direct relations between Eq.~($\ref{a21}$) and its sharp-interface counterpart, Eq.~($\ref{a18}$), we express $J_\rho(x)$, $J_e(x)$ and $\dot\eta$ in terms of $v$ and $J_T$. First, we make analysis considering only fluxes that are flowing through the free surface along $x$ direction (i.e normal to the free surface) thereby we tentatively drop the contribution of the surface diffusion flux $J_\text{sf}$ to $v$ since it is tangential to the free surface. Second, we employ a quasisteady approximation that assumes large gradients of $\rho$, $e$ and $\eta$ across the free surface such that we define their time derivatives as: 
\begin{equation}
\dot{\rho} \approx -v\rho^{'}(x), \: \dot{e} \approx -ve^{'}(x), \: \dot{\eta} \approx -v\eta^{'}(x).\label{approx}
\end{equation}

\noindent We integrate both sides of the conservation laws, $\dot e = -J_e^{'}(x)$ and $\dot \rho = -J_\rho^{'}(x)$ after substituting Eq. \eqref{approx};
\begin{equation}
    \int_{J_{\ss(\rho)}}^{J_{\at(\rho)}} \dif J_\rho =v\int^{\rho_\text{eq}^\ss}_{\rho_\text{eq}^\at} \dif \rho,\quad
    \int_{J_{\ss(e)}}^{J_{\at(e)}} \dif J_e =v\int_{e_\text{eq}^\ss}^{e_\text{eq}^\at}\dif e
    \label{intg}
\end{equation}
with the boundary values as
\begin{equation}
    \begin{split}
        J_{\ss(\rho)} \approx v\rho_\text{eq}^\ss,&\quad
        J_{\at(\rho)} \approx v\rho_\text{eq}^\at,\\
        J_{\ss(e)} \approx ve^\ss_\text{eq} - J_T, &\quad
        J_{\at(e)} \approx ve^\at_\text{eq} - J_T,
    \end{split}
\end{equation}
where $\rho^\text{eq}_\text{rg}$ and $e^\at_\text{rg}$ ($\text{rg} = \ss,~\at$) are the equilibrium conserved OP and internal energies, respectively. The integrals in Eq.(\ref{intg}) yield
\begin{equation}
 J_e(x) \approx ve(x) - J_T,  \quad J_\rho(x) \approx v\rho(x) \label{an14}.
\end{equation}
{Furthermore, we adopt the sigmoid formulation for the profiles of $\rho$(x) and $\eta(x)$ in this work as
\begin{align}
   \rho(x)&=\frac{1}{2}\left[\left(\rho_{\ss}+\rho_{\at}\right)+\left(\rho_{\ss}-\rho_{\at}\right)\tanh\frac{2x}{l}\right],
   \label{eq:rho}\\
   \eta(x) &= \frac{1}{2}\left[1 + \tanh\left(\frac{2x}{l}\right)\right]
\label{eq:sigmoid}
\end{align}
with the diffuse interface width $l$.} 
Taking into account all the aforementioned, we obtain $\sigma_s$ to be
\begin{eqnarray}
\sigma_s =&& \int^{l/2}_{-l/2}\bigg[\frac{(v\rho(x))^2}{{L}_{\rho\rho}} - \frac{(v\rho_\ss)^2}{2{L}^\ss_{\rho\rho}} - \frac{(v\rho_\at)^2}{2{L}^\at_{\rho\rho}}\bigg]\dif x   + \int^{l/2}_{-l/2}\bigg[\frac{(ve(x) - J_T)^2}{{L}_{ee}} - \frac{(ve_\ss - J_T)^2}{2{L}^\ss_{ee}} - \frac{(ve_\at - J_T)^2}{2{L}^\at_{ee}} \bigg]\dif x \nonumber \\ &&- \int^{l/2}_{-l/2} 4l\rho^{'}(x) \eta^{'}(x) v\bigg[M_{1}(v\rho(x)) + M_{2}(ve(x) - J_T)\bigg]\dif x \nonumber \\ &&+ \int^{l/2}_{-l/2} L_{\eta,\text{sf}}^{-1}v^2(\eta_{\text{eq}}^{'}(x))^2 \dif x. \label{a26}
\end{eqnarray}
It should be noted that the integration range of Eq.~(\ref{a26}) can also be taken from $-\infty$ and $+\infty$ without $\sigma_s$ changing. 
In this regard, we extend the integration interval from $[-l/2, +l/2]$ to $[-\infty, +\infty]$ in the following discussion. Comparing Eqs.~(\ref{a18}) and (\ref{a26}), we obtain:
\begin{eqnarray}
\mathcal{A} =&& \int^{\infty}_{-\infty} \bigg[\frac{\rho^2(x)}{{L}_{\rho\rho}} - \frac{(\rho_\ss)^2}{2{L}^\ss_{\rho\rho}} - \frac{(\rho_\at)^2}{2{L}^\at_{\rho\rho}} \bigg]\dif x - 4\int^{\infty}_{-\infty} M_1l\rho^{'}(x)\eta^{'}(x)\rho(x)\dif x \nonumber \\ &&+ \int^{\infty}_{-\infty} \bigg[ \frac{e^2(x)}{{L}_{ee}} - \frac{(e_\ss)^2}{2{L}^\ss_{ee}} - \frac{(e_\at)^2}{2{L}^\at_{ee}}\bigg]\dif x   -4\int^{\infty}_{-\infty} M_{2}l\rho^{'}(x)\eta^{'}(x) e(x)\dif x \nonumber \\ && + \int^{\infty}_{-\infty}  L_{\eta,\text{sf}}^{-1}[\eta^{'}(x)]^2\dif x, \label{a27}
\end{eqnarray}

\begin{equation}
\mathcal{B} = \int^{\infty}_{-\infty}2M_{2}l\rho^{'}(x)\eta^{'}(x)\dif x -\int^{\infty}_{-\infty} \bigg[ \frac{e(x)}{{L}_{ee}} - \frac{e^\ss}{2{L}^\ss_{ee}} - \frac{e^\at}{2{L}^\at_{ee}}\bigg]\dif x  , \label{a28}
\end{equation}

\begin{equation}
\mathcal{C} = \int^{\infty}_{-\infty} \bigg[\frac{1}{{L}_{ee}} - \frac{1}{2{L}^\ss_{ee}} - \frac{1}{2{L}^\at_{ee}} \bigg] \dif x,
\label{a29}
\end{equation}

\end{widetext}

\noindent The explicit formulations of $\mathcal{A}$, $\mathcal{B}$, and $\mathcal{C}$ implies that the phase-field parameters can be carefully tuned so as to obtain $\mathcal{A} = 0$, $\mathcal{B} = 0$ and $\mathcal{C} = 0$ which guarantees $\delta \mu = 0 $ and  $\delta T = 0$ across a migrating free surface. However, it is important to note that even if $\delta \mu = \delta T = 0$ is guaranteed, Almgren \cite{Almgren1999} showed that for phase-field models, conservation law at interfaces with opposing phases having asymmetry mobility coefficients has in existence two additional terms: interface stretching term and surface diffusion term. {In the sintering system, interface stretching represents excess mass and internal energy along the arclength of the free surfaces of the solid phase \cite{echebarria2004} and these excesses can both be respectively eliminated if $\int^{\infty}_{-\infty}\dif x[\rho_\text{eq} - \rho^\ss_\text{eq}/2 - \rho^\at_\text{eq}/2] = 0 $ and $\int^{\infty}_{-\infty}\dif x[e_\text{eq} - e^\ss_\text{eq}/2 - e^\at_\text{eq}/2] = 0 $ \cite{Almgren1999,Boussinot2014}. Taking $\rho_\text{eq}$ as defined in Eq.~\eqref{eq:rho} ensures that the interface excess of $\rho_\text{eq}$ is eliminated. Also, the interface excess of $e_\text{eq}$ is eliminated if $h(\rho)$ is taken as an odd function.} Furthermore, surface diffusion terms in the mass and energy conservation laws at the free surfaces of the solid are respectively parameterized by the mobilities $L^\mathrm{sf}_{\rho \rho} = \int^{\infty}_{-\infty}\dif x [L_{\rho \rho}(\rho) - L_{\rho \rho}^\ss/2 - L_{\rho \rho}^\at/2]$ and $L^\mathrm{sf}_\text{ee} = \int^{\infty}_{-\infty}\dif x [L_\text{ee}(\rho) - L_\text{ee}^\ss/2 - L_\text{ee}^\at/2]$. \cite{Almgren1999,Boussinot2014}. 

In order to make $\delta T = 0$, we need to ensure that $\mathcal{B} = \mathcal{C} = 0$. Consequently, $L_\text{ee}$ should be formulated such that it gives the bulk region energy mobilities at the corresponding regions, ensures $\mathcal{C}=0$ and also guarantee that the model replicates the sharp-interface energy conservation law (Eq. \eqref{si5}) where there is no surface diffusion effect (i.e $L^\mathrm{sf}_\text{ee} = 0$). To achieve this, Almgren \cite{Almgren1999} proposed a mobility interpolation function which is a combination of odd functions with parameters adjusted relative to the bulk mobilities. This method is contended by Ohno et al. \cite{ohno2016} as the mobility interpolation function produces a non-monotonic function and also contributes to a limited ratio of the possible bulk mobilities. Nevertheless, it is vital to note that while simultaneous elimination of $\delta T$ and surface diffusion effect somewhat put constraints on a scalar formulation of $L_\text{ee}$, the emergence of both effects is actually direction dependent \cite{nicoli2011}. $L_\text{ee}$ formulation constraint due to $\delta T$ (Eq.~(\ref{a29})) emerges under the consideration of flux components normal to the free surfaces as seen in the analysis done above while the integral associated with the surface diffusion effect modification of energy conservation emanates due to consideration of flux components in tangential direction to the free surfaces \cite{nicoli2011}. Therefore, ensuring $\mathcal{C}=0$ and eliminating surface diffusion term in energy conservation equation are respectively pertinent only at the normal and tangential directions of the free surfaces. Considering all the aforementioned and also taking into account the physical context of the energy mobility, we propose an anisotropic $\mathbf{L}_{ee}$ for the full sintering description and relate it to the anisotropic thermal conductivity as

\begin{equation}
\begin{split}
\mathbf{L}_\text{ee} &= \left[k_{\perp}\mathbf{N}_\mathrm{sf} + k_{\parallel}\mathbf{T}_\text{sf} + k_\text{gb}\mathbf{T}_\text{gb}\right]T^2 \\
&=L^{\perp}_\text{ee}\mathbf{N}_\mathrm{sf} + L^{\parallel}_\text{ee}\mathbf{T}_\text{sf} + L^\text{gb}_\text{ee}\mathbf{T}_\text{gb}, \label{an8}
\end{split}
\end{equation}
with
\begin{equation}
k_{\perp}= \left[\frac{1+g(\rho)}{2{k}_\ss} + \frac{1-g(\rho)}{2{k}_\at}\right]^{-1}, \label{an9}
\end{equation}
\begin{equation}
k_{\parallel} = \frac{1+g(\rho)}{2}k_\ss + \frac{1-g(\rho)}{2}k_\at, \label{an10}
\end{equation}
\begin{equation}
k_\text{gb}=16\sum_{i\neq j}\eta_i^2\eta_j^2k_\text{gb},
\end{equation}
and 
\begin{equation}
\begin{aligned}
\mathbf{N}_\mathrm{sf}&= \mathbf{n}_\text{sf}\otimes\mathbf{n}_\text{sf},\\ \mathbf{T}_\mathrm{sf}&=\mathbf{I} - \mathbf{n}_\text{sf}\otimes\mathbf{n}_\text{sf},
\\
\mathbf{T}_\mathrm{gb}&=\mathbf{I} - \mathbf{n}_\text{gb}\otimes\mathbf{n}_\text{gb}.
\end{aligned}
\end{equation}
In Eq~\eqref{an8}, $L^{\perp}_\text{ee}$ is the energy mobility in normal direction to the free surfaces defined to ensure $\mathcal{C}=0$, $L^{\parallel}_\text{ee}$ is the energy mobility in the tangential direction to the free surfaces formulated to ensure $L^\mathrm{sf}_\text{ee}=0$ in the energy conservation law, and $L^\text{gb}_\text{ee}$ represents the energy mobility in the grain boundary. Similarly, $k_{\perp}$ and $k_{\parallel}$ represent the thermal conductivities at the normal and tangential directions to the free surfaces respectively while $k_\text{gb}$ represent the thermal conductivity in the grain boundary.
${k}_\ss$ and ${k}_\at$ are respectively the effective thermal conductivities in the solid phase and atmosphere region and $k_\text{gb}$ is the effective thermal conductivity in the grain boundary. Surface and grain boundary normal vectors are calculated from the gradient of corresponding OPs, e.g., $\mathbf{n}_\text{sf}\equiv\nabla\rho / |\nabla\rho|$. $\mathbf{I}$ is the identity tensor and $\otimes$ represents the dyadic product. $g(\rho) = 2\rho - 1$ is an odd function that satisfies $g(\rho=\rho_\ss) = 1$ and $g(\rho=\rho_\at)= -1$. 

Noting that $L^{\perp}_\text{ee} = k_{\perp}T^2$ and therefore substituting Eq.~($\ref{an9}$) into ($\ref{a28}$), we obtain
\begin{equation}
\mathcal{B} = 2\chi M_2 - \frac{\beta l}{2T^2}\left( \frac{1}{2{k}_\ss} - \frac{1}{2{k}_\at} \right),
\end{equation}
{with
\begin{equation}
\chi = l\int^{\infty}_{-\infty}\rho^{'}(x)\eta^{'}(x) \dif x = 2(\rho_{\ss} - \rho_{\at})/3, \label{eq:integral1}
\end{equation}
\begin{equation}
\beta = \frac{e_\text{ht}}{l}\int^{\infty}_{-\infty}[h(\rho)g(\rho)-1]  \dif x = -e_\text{ht}(\rho_{\ss} - \rho_{\at})^2, \label{eq:integral2}
\end{equation}}
\noindent where $e_\text{ht} =e_\ss - e_\at$. The functions defined in Eq. \eqref{eq:sigmoid} and \eqref{eq:rho} is adopted to calculate integrals in Eqs. \eqref{eq:integral1} and \eqref{eq:integral2}.

\noindent Therefore, to obtain $\mathcal{B} =0$, we take
{\begin{equation}
M_2 = \frac{\beta l}{4\chi T^2}\left( \frac{1}{2{k}_\ss} - \frac{1}{2{k}_\at} \right). \label{an15}
\end{equation}}

Following \cite{ahmed2013,yangscripta2020,tonks2015development} whereby the different mass diffusion routes in sintering process i.e bulk/volume diffusion, surface diffusion along the free surfaces and grain boundary diffusion are taking into account, we propose an anisotropic $\mathbf{L}_{\rho\rho}$ and relate it to the anisotropic diffusivity as
\begin{equation}
\begin{split}
\mathbf{L}_{\rho\rho}&=  \left[D_\mathrm{v}\mathbf{I} + D_\mathrm{sf}\mathbf{T}_\text{sf} + D_\text{gb}\mathbf{T}_\text{gb}\right]/s_{\mathrm{v}} \\
&=L^\mathrm{v}_{\rho\rho}\mathbf{I} + L^\mathrm{sf}_{\rho\rho}\mathbf{T}_\text{sf} + L^\text{gb}_{\rho\rho}\mathbf{T}_\text{gb}, \label{an11}
\end{split}
\end{equation}
with
\begin{equation}
D_\mathrm{v} = \left[\frac{1+g(\rho)}{2{D}_\ss} + \frac{1-g(\rho)}{2{D}_\at}\right]^{-1}, \label{an13n}
\end{equation}
\begin{equation}
D_\mathrm{sf} =16\rho^2(1-\rho)^2 D_\text{sf},
\end{equation}
\begin{equation}
D_\text{gb} =16\sum_{i\neq j}\eta_i^2\eta_j^2 D_\text{gb},
\end{equation}
and the volumetric entropy as
\begin{equation}
s_\mathrm{v} = \frac{1}{T}\frac{\partial\mu}{\partial \rho}, 
\end{equation}
where the linear approximation is sometimes taken as $s_\mathrm{v}\approx \mathcal{R}/V_\mathrm{m}$ with the ideal gas constant $\mathcal{R}$ and molar volume $V_\mathrm{m}$ \cite{ahmed2013,zhang2012phase,yangscripta2020}. In Eq.~\eqref{an11}, $L^\mathrm{v}_{\rho\rho}$ is the mass mobility in the normal direction to the free surfaces associated with the bulk/volume diffusion in solid phase and atmosphere region, $L^\mathrm{sf}_{\rho\rho}$ is the mass mobility in the tangential direction to the free surfaces associated with mass transport via surface diffusion. {Consideration of $L^\mathrm{sf}_{\rho\rho}$ ensures that the model replicates the sharp-interface mass conservation law Eq. \eqref{sinw} where surface diffusion is considered}. $L^\text{gb}_{\rho\rho}$ represent the mass mobility in the grain boundary. Similarly, $D_\mathrm{v}$ represents the volume diffusivity, which is interpolated by the effective diffusivities in the solid phase (${D}_\ss$) and atmosphere region (${D}_\at$). $D_\text{sf}$ and $D_\text{gb}$ are the effective diffusivities in the free surfaces and grain boundary, respectively.

We propose $M_1$ to have a similar formulation as $M_2$ in Eq. \eqref{an15};
{\begin{equation}
M_1 = - \frac{3ls_\mathrm{v}}{16}\left(\frac{A_\mathrm{ss}}{2{D}_\ss} - \frac{A_\mathrm{at}}{2{D}_\at}\right) \label{an16},
\end{equation}
with
\begin{align}
   A_\mathrm{ss} &= \rho_{\ss}+\rho_{\at}, \\
   A_\mathrm{at} &= 3A_\mathrm{ss}-2. 
\end{align}}

Substituting Eqs.~$\eqref{an15}$ and $\eqref{an16}$ into $\eqref{a27}$, we then obtain
\begin{equation}
\mathcal{A} = \frac{\psi L_{\eta,\text{sf}}^{-1}}{l} - \frac{l(\rho_{\ss}-\rho_{\at})^2}{4} \left[\frac{s_\mathrm{v}A_\mathrm{ss}}{{D}_\ss} + \frac{\zeta }{T^2}\left(\frac{1}{2{k}_\ss} + \frac{1}{2{k}_\at} \right)\right],
\end{equation}
with 
\begin{equation}
\zeta = \frac{e^2_\text{ht}}{l}\int^{\infty}_{-\infty}[1-h^2(\rho)] \dif x = e^2_\text{ht}, \label{eq:integral6}
\end{equation}
and
\begin{equation}
\psi =  l\int^{\infty}_{-\infty}(\eta^{'}(x))^2
\dif x = 2/3, \label{eq:integral3}
\end{equation}
also taking into account:
\begin{equation}
\begin{split}
\int^{\infty}_{-\infty} \left[(\rho_\text{eq}(x))^2 + g(\rho_\text{eq})(\rho_\text{eq}(x))^2 -1\right] \dif x  \\ = -\frac{3l}{4}A_\mathrm{ss}(\rho_{\ss}-\rho_{\at})^2,
\end{split}
\end{equation}

\begin{equation}
\int^{\infty}_{-\infty} \left[(\rho_\text{eq}(x))^2[1 - g(\rho_\text{eq})\right] \dif x = \frac{l}{4}A_\mathrm{at}(\rho_{\ss}-\rho_{\at})^2.
\label{eq:integral5}
\end{equation}
The functions defined in Eq. \eqref{eq:sigmoid} and \eqref{eq:rho} are again adopted to calculate integrals in Eq. \eqref{eq:integral3} and Eqs. \eqref{eq:integral6} - \eqref{eq:integral5}.

\noindent Therefore, in order to ensure $\mathcal{A}=0$, we take
\begin{equation}
L_{\eta,\text{sf}}^{-1} = \frac{l^2(\rho_{\ss}-\rho_{\at})^2}{4\psi} \left[\frac{s_\mathrm{v}A_\mathrm{ss}}{{D}_\ss} + \frac{\zeta }{T^2}\left(\frac{1}{2{k}_\ss} + \frac{1}{2{k}_\at} \right)\right].
\end{equation}

\noindent{The mobility $L_{\eta,\text{gb}}^{-1}$ of $\{\eta_i\}$ can be obtained from the physical grain boundary mobility $G_\text{gb}^\text{eff}$ and grain boundary energy $\gamma_\text{gb}$ as \cite{moelans2008, yangscripta2020} 
\begin{equation}
L_{\eta,\text{gb}}^{-1} = \frac{\kappa_{\eta}}{G_\text{gb}^\text{eff} \gamma_\text{gb}}.
\end{equation}
Recalling Eqs. \eqref{an6} and \eqref{an7}, it is worth mentioning that this mobility is defined under the driving force represented by entropy, which should be distinguished from the original formulation in Ref. \cite{moelans2008} as here $\kappa_\eta$ adopts the dimension of the entropy per length.} 
Accordingly, for $L_{\eta}^{-1}$ as regards the full sintering description, we then take
\begin{equation}
L_{\eta}^{-1} = 16\rho^2(1-\rho)^2 L_{\eta,\text{sf}}^{-1} +  L_{\eta,\text{gb}}^{-1}.
\end{equation}
Recalling the anisotropic definitions of $\mathbf{L}_{ee}$ and $\mathbf{L}_{\rho\rho}$ in Eqs. \eqref{an8} and \eqref{an11}, calculation of $\hat{L}_\eta^{-1}$ in Eq. \eqref{an7} can be further simplified as
\begin{equation}
    \hat L_\eta^{-1}=L_\eta^{-1} - l^2|\nabla\rho|^2[M^2_1 L_{\rho \rho}^\mathrm{v}  + M_2^2 L_{ee}^\perp],
\end{equation}
as $L_{ee}^\perp$ and $L_{\rho \rho}^\mathrm{v}$ are respectively one of the eigen-values of $\mathbf{L}_{ee}$ and $\mathbf{L}_{\rho\rho}$,  corresponding to the eigen-direction of $\mathbf{n}_\mathsf{sf}$ ($\mathbf{n}_\mathsf{sf}\equiv\nabla\rho/|\nabla\rho|$).

It is worth noting that $M_1$ and $M_2$ are derived based on the constant postulate, i.e., $M_1$ and $M_2$ are spatio-temporal independent constants for a sintering system with known mass diffusivities and thermal conductivities of substance and atmosphere as well as given diffuse interface width, since the spatio-temporal dependency of all OP-related terms (Eqs. \eqref{eq:integral1}-\eqref{eq:integral2}, \eqref{eq:integral3}, and \eqref{eq:integral6}-\eqref{eq:integral5}) vanish after integral. 
More importantly, the quantitative phase-field model degenerates to the conventional one when the system has no 
differences in mass diffusivity and thermal conductivity between solid and atmosphere. In that sense, when ${D}_\ss = {D}_\at$, $M_1 = 0$ and also when ${k}_\ss = {k}_\at$, $M_2 = 0$, demonstrating that 
the antitrapping terms in Eqs \eqref{ann4a} and \eqref{an5} and cross-coupling term in Eq. \eqref{an6} reduce to zero. 
{In addition, we note that variational quantitative phase-field models such as the one presented in this work do not generally demonstrate high numerical accuracy \cite{karma1998, ohno2016, ohno2017}. Correct mapping of the variational model onto the associated sharp-interface equations only guarantees its quantitative validity and not its numerical efficiency needed for realistic utilization \cite{ohno2016}. Therefore, a nonvariational form of the model might be best suited for practicability. The nonvariational form can be simply developed via modification of model parameters and functions while ensuring that the thin-interface asymptotic remains consistent.}

\begin{table*}
\centering
\caption{\label{table1} Set of dimensionless quantities and parameters employed for the simulations in this work.} \setlength{\tabcolsep}{6mm}{
\begin{tabular}{llllllll}
\toprule
         $L_x$   & $L_y$ & $g_0$ & $\underline{C}$ & $\underline{D}$ & $e_\text{ht}$ & $L_{\eta,\text{gb}}$ & $s_\mathrm{v}$\\
            \hline
60   &  50   &   0.01 &  1   &  0.062 & 1 & 1 & 1\\
\toprule
\end{tabular}}
\end{table*}

\begin{figure*}
\includegraphics[width=16cm]{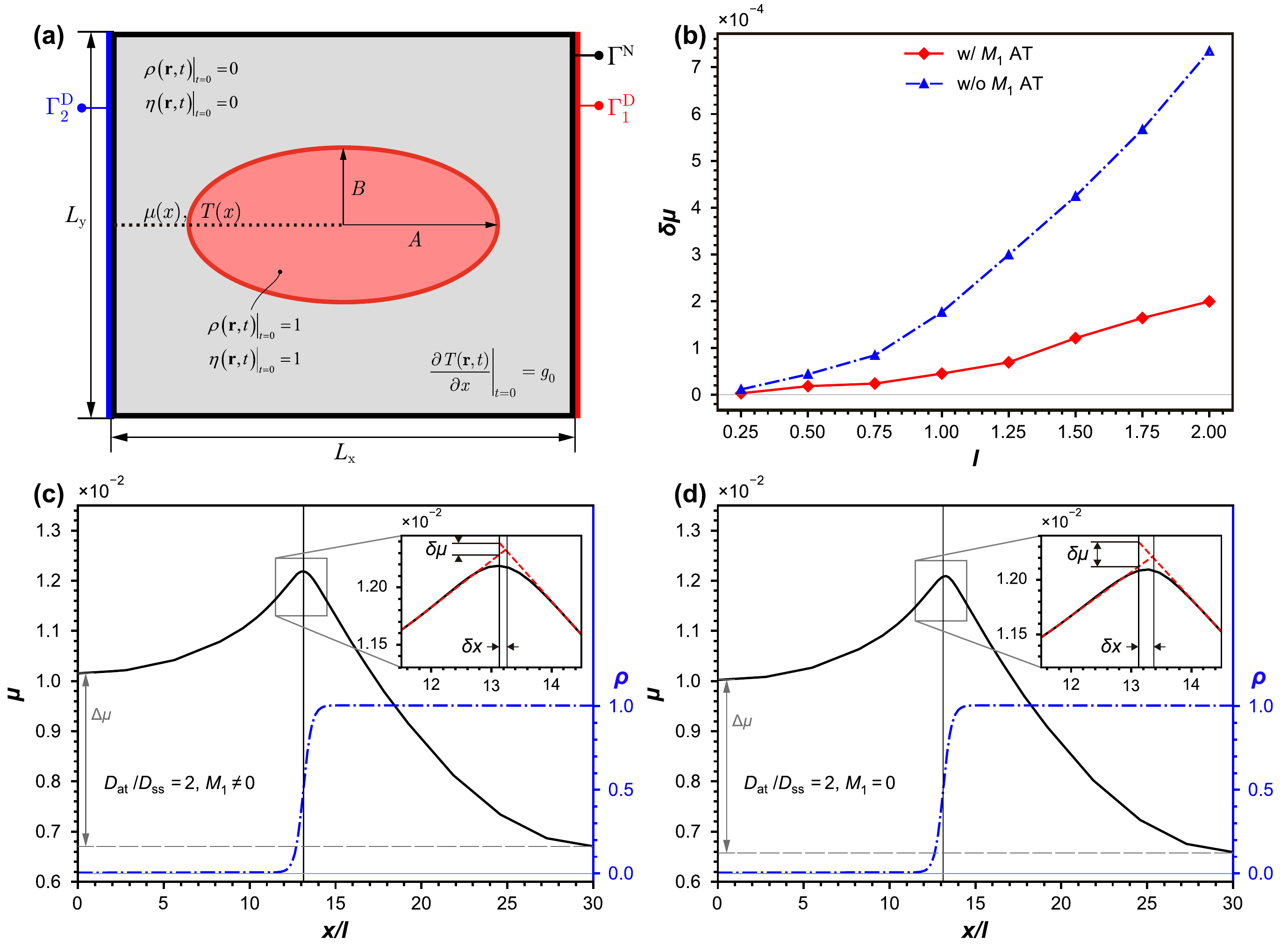}
\caption{\label{fig:ellipse} (a) Schematic of the simulation setup in 2D simulation of an elliptical inclusion (b) Comparison of chemical potential jump $\delta \mu$ across the free surface with respect to interface width; blue triangle symbols represent model with $M_1 = 0$ while red diamond symbols represent model with $M_1 \neq 0$. Plots of $\mu$ and $\rho$ across the free surface as a function of $x$ with $l = 1$, $D_\at/D_\ss=2$  for (c) $M_1 \neq 0$ and (d) $M_1 = 0$. $\delta \mu$ is obtained using an extrapolation of $\mu$ at the center of the free surface $\rho = 0.5$. $\Delta \mu$ is the chemical potential difference between the bulk values.} 
\end{figure*}

\section{\label{sec:results}Results and discussion}
\subsection{Model verification for an elliptical inclusion}
In order to examine the capability of the model in ensuring $\delta \mu = \delta T = 0$ at the free surface, we perform diffusion-driven reshaping simulations of an elliptical inclusion with major axis $A$ and minor axis $B$ morphing into a circle. We set up a simulation domain with the lengths $L_x$ and $L_y$ in $x$ and $y$ direction, respectively. The  domain is further subjected to an initial temperature gradient $\nabla T = g_0$ along the x-axis. {$\rho$ is taken to vary smoothly from one in the inclusion ($\rho_{\ss} = 1$) to zero outside ($\rho_{\at} = 0$) with $l = 1$}. A full schematic of the simulation setup is given in Fig.~\ref{fig:ellipse}a.  {The normalized values of the employed model parameters are given in Table~\ref{table1}.} 

First, we consider a case of asymmetric mass transport where $D_\at/D_\ss=2$. We set $k_\at/k_\ss=1$, hence only employing the mass antitrapping current term associated with mass diffusion while the thermal antitrapping current is tentatively dropped. Profile of chemical potential $\mu (x)$ across the moving free surface is presented for the cases $M_1 = 0$ and $M_1 \neq 0$ in Figs.~\ref{fig:ellipse}c and \ref{fig:ellipse}d, respectively. An extrapolation of $\mu(x)$ gives the chemical potential jump ($\delta \mu$) at the center of the free surface $\rho =0.5$. Typically, $\delta \mu \neq 0$ implies an exchange of mass between the solid and atmosphere, which can be likened to the trans-interface diffusion phenomenon. However, no mass exchange is expected between the solid and atmosphere regions during sintering. Therefore, $\delta \mu = 0$ should be held in phase-field simulations in order to achieve realistic mass diffusion. It is obvious from Figs.~\ref{fig:ellipse}c and \ref{fig:ellipse}d that the case with $M_1 = 0$ shows a significantly larger $\delta \mu$ compared to the one with $M_1 \neq 0$, in which the relatively small $\delta \mu$ is attributed to possible numerical errors. The results demonstrate that the mass antitrapping current parameterized by $M_1$ is necessary in order to eliminate the artificial diffusion flux across the interface during sintering for cases of asymmetric mass transport. Figs.~\ref{fig:ellipse}c and \ref{fig:ellipse}d also show the gap in space $\delta x$ between the center of the free surface and the point where the extrapolations of $\mu$ meet. Note that $\delta x = 0$ when $\delta \mu = 0$, indicating the coherence between the numerically predicted interface by $\rho \approx 0.5$ and the theoretical sharp interface where $\delta \mu = 0$. Similar to $\delta \mu$, the numerical results demonstrate a significantly larger $\delta x$ for the case with $M_1 = 0$ compared to the one with $M_1 \neq 0$, implying an apparent deviation in the position between the predicted interface and theoretical sharp interface. 

{We also note the existence of another chemical potential drop $\Delta \mu$ across the free surface, characterizing the differences between the bulk values, as depicted in Figs.~\ref{fig:ellipse}c and \ref{fig:ellipse}d. This $\Delta \mu$, which is identical for both cases at a time point, were numerically examined to be the outcome of the deviated bulk values of $\rho$, i.e., $\rho_\ss$ and $\rho_\at$ that are slightly deviated from ideal (equilibrium) one and zero respectively, as listed in Table. S1.
Such chemical potential drop generally does not appear in the conventional sharp-interface interpretation of the sintering \cite{german2014sintering, kang2004}. Meanwhile, the deviated bulk values of the conserved mass OP have been depicted in previous works \cite{jacqmin1999, yue2004, feng2005, yue2007, dadvand2021} with theoretical and numerical analyses given in Refs. \cite{yue2007} and \cite{dadvand2021}, which are further discussed in the Appendix.}


Furthermore, Fig.~\ref{fig:ellipse}b shows a comparison of $\delta \mu$ vs. diffuse interface width $l$ between the cases with/without $M_1$ parameterized. It can be observed that both cases present the convergence $\delta \mu  \to 0$ as $l \to 0$, replicating the sharp-interface condition when $l$ tends to infinitesimal. However, as $l$ increases, $\delta \mu$ present a relatively rapid growth in the case with $M_1 = 0$ compared to the one with $M_1 \neq 0$, demonstrating that the employment of mass antitrapping current parameterized by $M_1$ can significantly reduce the artificial interface effect (here the growing $\delta \mu$) along with increasing diffuse interface width. In this sense, mass antitrapping current allows reasonable quantitative simulations, especially at larger interface widths. Furthermore, we note that the convergence of both models might be well investigated considering a steady state free surface velocity. This we hope to report in our upcoming work. 

Additionally, we investigate a case of asymmetric heat transport with $k_\at/k_\ss=0.05$. Similar to previous simulation, we examine the thermal antitrapping term associated with heat transport. The mass antitrapping term is tentatively dropped by setting $D_\at/D_\ss=1$.  Simulations are performed for existing model (i.e., $M_2 = 0$) and quantitative model with $M_2 \neq 0$. Further details of results are given in supplementary material (Fig. S2). $\delta T = 0$ realized at the sharp-interface is expected to be obtained during phase-field simulations in order to guarantee quantitative simulations. For model with $M_2 = 0$, however, emerging $\delta T \neq 0$ demonstrates the importance of the thermal antitrapping current. Here, it is important to note that measured $\delta T$ has a relatively low magnitude compared to the bulk temperature at the free surface. Importance of thermal antitrapping term $M_2$ in eliminating temperature jump for asymmetric heat transport has also been demonstrated in Ref. \cite{Boussinot2017} where non-diagonal phase-field model was also used.


\begin{figure*}
\includegraphics[width=14cm]{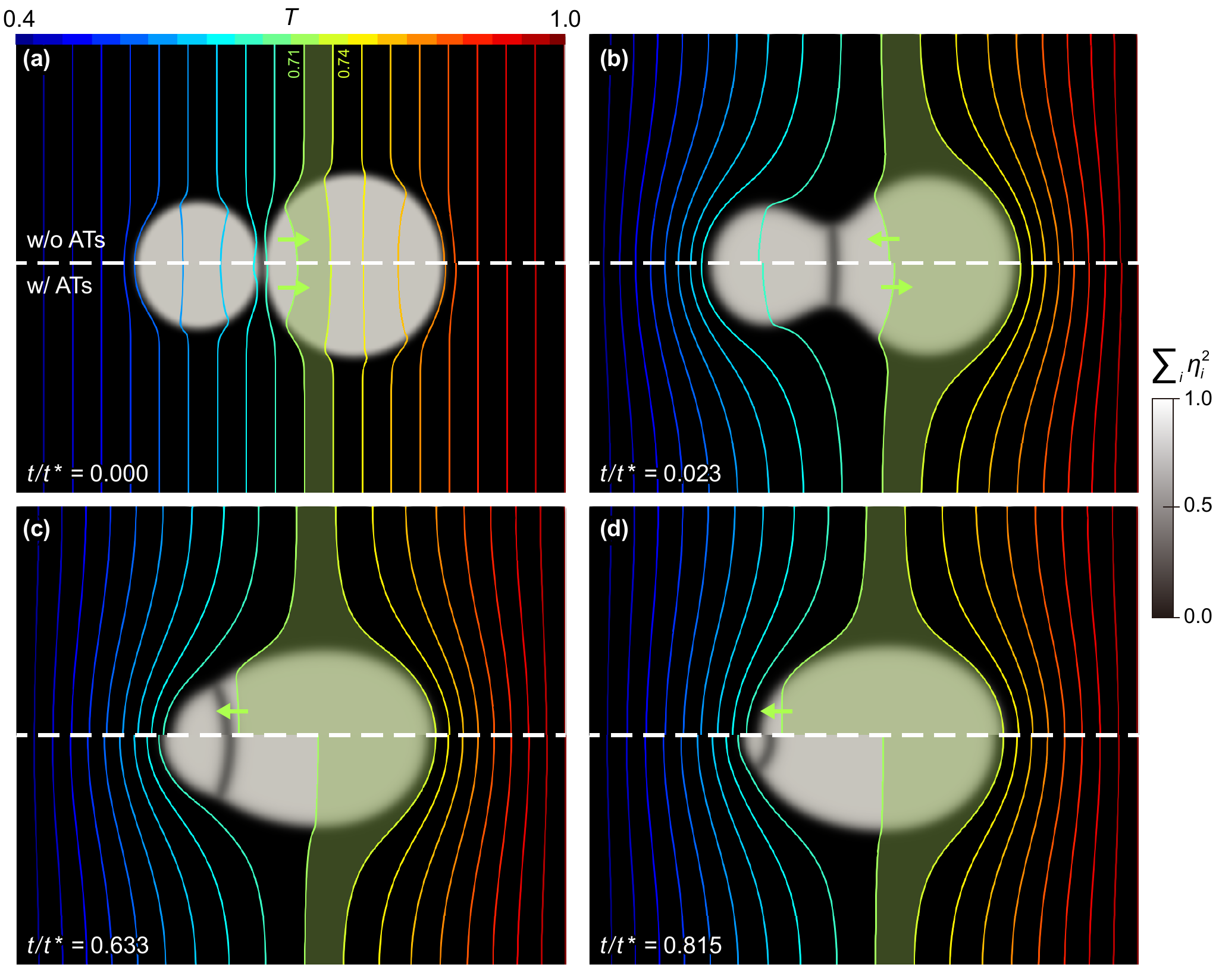}
\caption{\label{fig:comparison} Snapshots of grains coalescence of two spherical grains with distinct sizes. $D_\at/D_\ss=2$, $l = 2$ and $k_\at/k_\ss=0.05$ are set. Comparison is made between models with ATs and without ATs. Temperature isolines are also indicated. $t^* = 10^3$ unit.} 
\end{figure*}

\begin{figure}
\includegraphics[width=8.5cm]{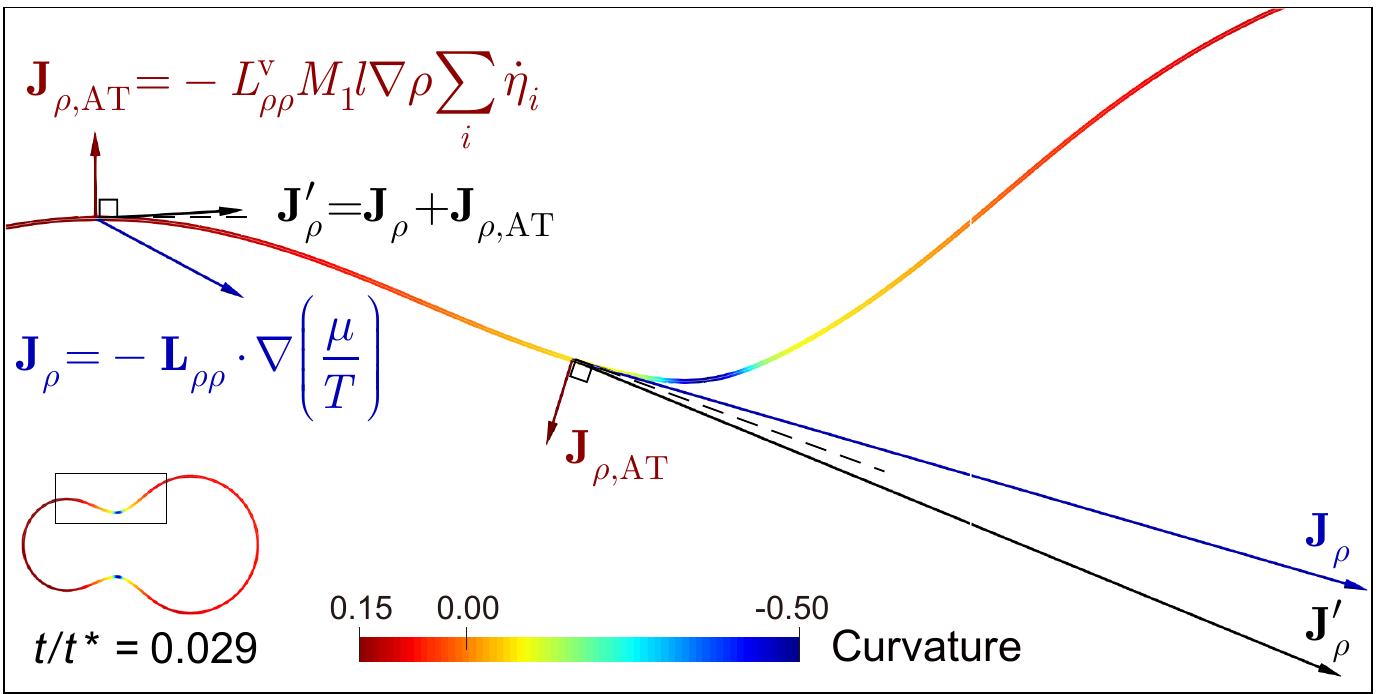}
\caption{\label{fig:curve} Surface profile ($\rho=0.5$) colored by the curvature and the mass diffusion fluxes, i.e., the fluxes before ($\mathbf{J}_\mathrm{\rho}$) and after ($\mathbf{J}_\mathrm{\rho}'$) correction with the antitrapping contribution ($\mathbf{J}_\mathrm{\rho,AT}$) , at two distinct sites. The length of visualized arrows have been scaled according to the magnitude of the fluxes uniformly.} 
\end{figure}

\subsection{Comparison between the quantitative and the existing models}
In this section, we perform simulations for grain coalescence of two spherical grains with distinct sizes. Comparisons of microstructure and temperature distribution are made between quantitative model where antitrapping terms are taken into account and existing model where these terms are not considered. The two models are referred to as model with ATs and model without ATs in the following discussions.

We set up a simulation domain with the lengths $L_x$ and $L_y$ in $x$ and $y$ direction, respectively. Similar to previous setup, the domain is subjected to $\nabla T = g_0$ along the x-axis. Simulations are performed for asymmetric mass and heat transport where $D_\at/D_\ss=2$, $k_\at/k_\ss=0.05$ with $l = 2$. A full schematic of the simulation setup is supplemented in Fig. S4a. 

Transient microstructures and temperature profiles for both models are compared and presented in Fig.~\ref{fig:comparison}. First, we observe that mass transport was faster for model without ATs compared to model with ATs. At $t / t^* = 0.633$ and $t / t^* = 0.815$ in Figs.~\ref{fig:comparison}c and \ref{fig:comparison}d respectively, a more coalesced grain is obtained for model without ATs compared to model with ATs. The difference in progress of coalescence can be further explained by the visualization of mass diffusion fluxes at the free surface $\rho = 0.5$ as presented in Fig.~\ref{fig:curve}. The free surface profile is colored by the local curvature calculated as $-\nabla \cdot \mathbf{n}_\mathrm{sf}$. Furthermore, mass diffusion fluxes are indicated at two distinct points; a concave point and a convex point. Typically, mass flux at any point on the free surface is expected to be correctly captured along the tangential direction to the free surface at that point. It can be clearly observed that $\mathbf{J}_\rho$ which is the mass diffusion flux without the mass antitrapping current deviates in direction from the tangential direction (dashdot lines) to the free surface at both concave and convex points. The mass antitrapping flux $\mathbf{J}_{\rho, AT}$ introduced in quantitative model can be seen flowing through the free surface in the normal direction from the solid grain region to the atmosphere. The combined mass flux $ \mathbf{J}^{'}_{\rho} = \mathbf{J}_{\rho} + \mathbf{J}_{\rho, AT}$ shows a corrected mass flux flowing along the tangential direction to the free surface. Therefore, the deviation of $\mathbf{J}_{\rho}$ from its appropriate direction is due to the existence of chemical potential jump at the free surface. $\mathbf{J}_{\rho, AT}$ serves to eliminate this chemical potential jump which consequently corrects this deviation. Accordingly, this demonstrates the faster mass transport observed for model without ATs. Chemical potential jump at the free surface tend to act as an extra driving force for grain coalescence leading to faster mass diffusion. The elimination of this jump via the antitrapping current leads to a slower mass transport for model with ATs. 

\begin{figure*}
\includegraphics[width=16cm]{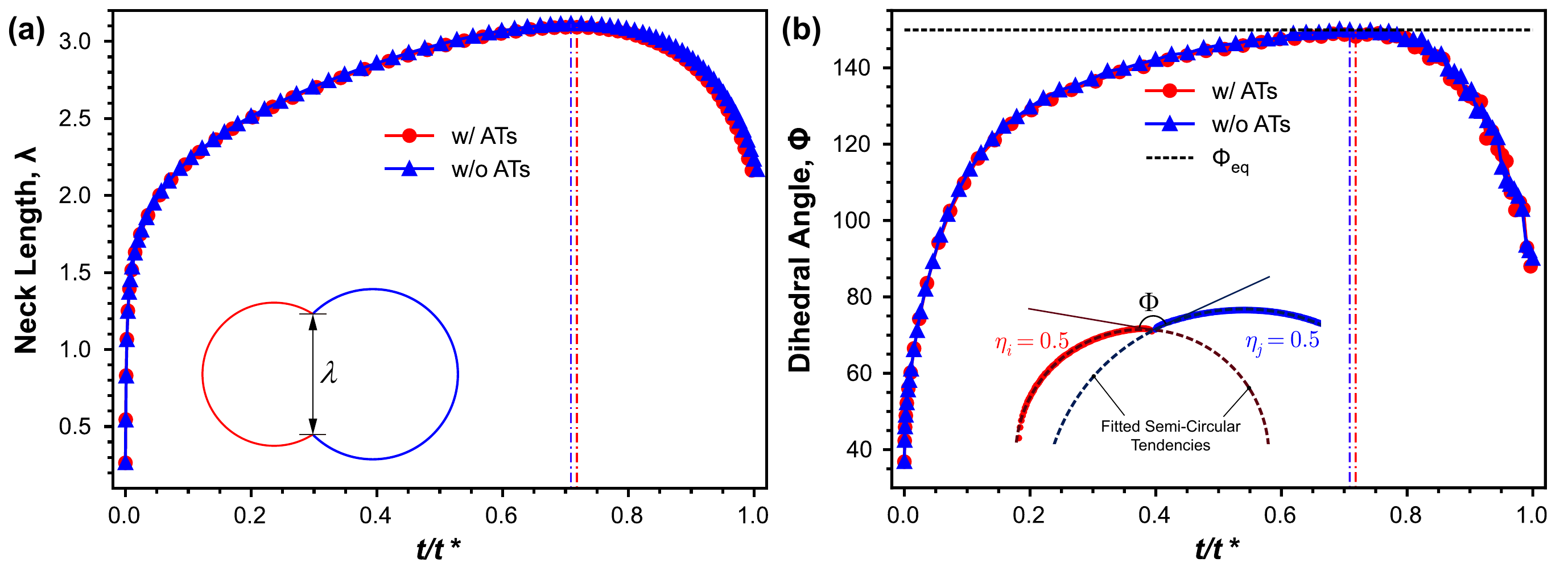}
\caption{\label{fig:neck_da} Time evolution of (a) sintering neck $\lambda$ and (b) dihedral angle $\Phi$ during the non-isothermal sintering process, as shown in Fig.~\ref{fig:comparison}. The time points reaching maximum values are indicated by colored vertical lines. The equilibrium dihedral angle $\Phi_\mathrm{eq}$, calculated from the surface and grain boundary energies, is also indicated by black dotted line in (b). The total simulation time $t^{*} = 2084$ unit.} 
\end{figure*}

Furthermore, results in Fig.~\ref{fig:comparison} also show the comparison of the temperature profiles obtained for both models. The distribution of the temperature isolines shows faster heat transport for model without ATs compared to model with ATs. An example is given for isoline $T = 0.71$. Even though it initially tends to migrate towards high-$T$ side, at $t / t^* = 0.023$, this tendency breaks for model without ATs, where the isoline starts to move towards low-$T$ side, but continues for model with ATs. Result at $t / t^* = 0.815$ in Fig.~\ref{fig:comparison}d indicates a colder grain for model with ATs compared to model without ATs. Similar to the mass transport fluxes explanation for both models, the temperature jump at the free surface can be seen as an extra driving force for heat transport in the model without ATs. This jump is eliminated for model with ATs via the thermal antitrapping term thereby obtaining a slower heat transport. The presented thermal-microstructure evolution once more demonstrates the importance of the antitrapping currents for mass and heat diffusion.

{We further examine the in-process sintering neck $\lambda$ and dihedral angle $\Phi$ of the simulation as presented in Fig. \ref{fig:comparison}. $\lambda$ and $\Phi$ are calculated by
\begin{equation}
\begin{aligned}
\lambda&=\int_\Omega \frac{16\sum_{i\neq j} \eta_i^2 \eta_j^2}{l}\dif\Omega, \\
\Phi&=\arctan\left(\frac{\partial C_{\eta_i} }{\partial x}\right)_\mathrm{neck}-\arctan\left(\frac{\partial C_{\eta_j} }{\partial x}\right)_\mathrm{neck},
\end{aligned}
\label{eq:neck_da}
\end{equation}
where $C_{\eta_i}$ and $C_{\eta_j}$ are the fitted semi-circular tendencies by coordinates of contour $\eta_i=0.5$ and $\eta_j=0.5$, respectively. $\frac{\partial C_{\eta_i}}{\partial x}$ and $\frac{\partial C_{\eta_j}}{\partial x}$ then provide the slopes of $C_{\eta_i}$ and $C_{\eta_j}$. In this sense, $\Phi$ is calculated using the difference between these two angles of slope at the neck point, as shown in inset of Fig.~\ref{fig:neck_da}b, adapted from Ref.~\cite{moelans2009}. Meanwhile, the equilibrium dihedral angle $\Phi_\mathrm{eq}$ can be also evaluated by the surface ($\gamma_\mathrm{sf}$) and grain boundary ($\gamma_\mathrm{gb}$) energies, i.e.,
\begin{equation}
\Phi_\mathrm{eq}=2\arctan\frac{\gamma_\mathrm{gb}}{2\gamma_\mathrm{sf}}.
\label{eq:da_theo}
\end{equation}
It is worth noting that $\Phi$ approaches $\Phi_\mathrm{eq}$ when two particles with identical size are sintered isothermally, as $\lambda$ reaches the maximum and stays constant, i.e., the system reaches equilibrium \cite{yangscripta2020}. With varying interface width $l$, $\Phi$ deviates from theoretically-determined $\Phi_\mathrm{eq}$ (Eq. \eqref{eq:da_theo}) as shown in Fig. S3. This deviation is reduced in a similar fashion for both models with/without ATs as $l$ decreases. This implies no modification to thermodynamic equilibrium condition (characterized by $\Phi_\mathrm{eq}$) by applying the kinetic antitrapping terms.
For two non-identical grains, the time evolution of $\lambda$ and $\Phi$ are presented in Fig.~\ref{fig:neck_da}. Comparison is made for model with ATs and model without ATs. It can be observed that for both models, $\Phi$ approaches $\Phi_\mathrm{eq}$ at the points where $\lambda$ attains maximum values. However, the progress of $\Phi$ towards $\Phi_\mathrm{eq}$ is faster for model without ATs compared to model with ATs. This implies that while thermodynamic conditions are attained for both models, the antitrapping terms tend to modify the progress of neck growth and grain coalescence by removing the extra flux perpendicular to the free surface, as evidently shown in Fig.~\ref{fig:comparison} and Fig. \ref{fig:curve}. 
}

\begin{figure*}
\includegraphics[width=14cm]{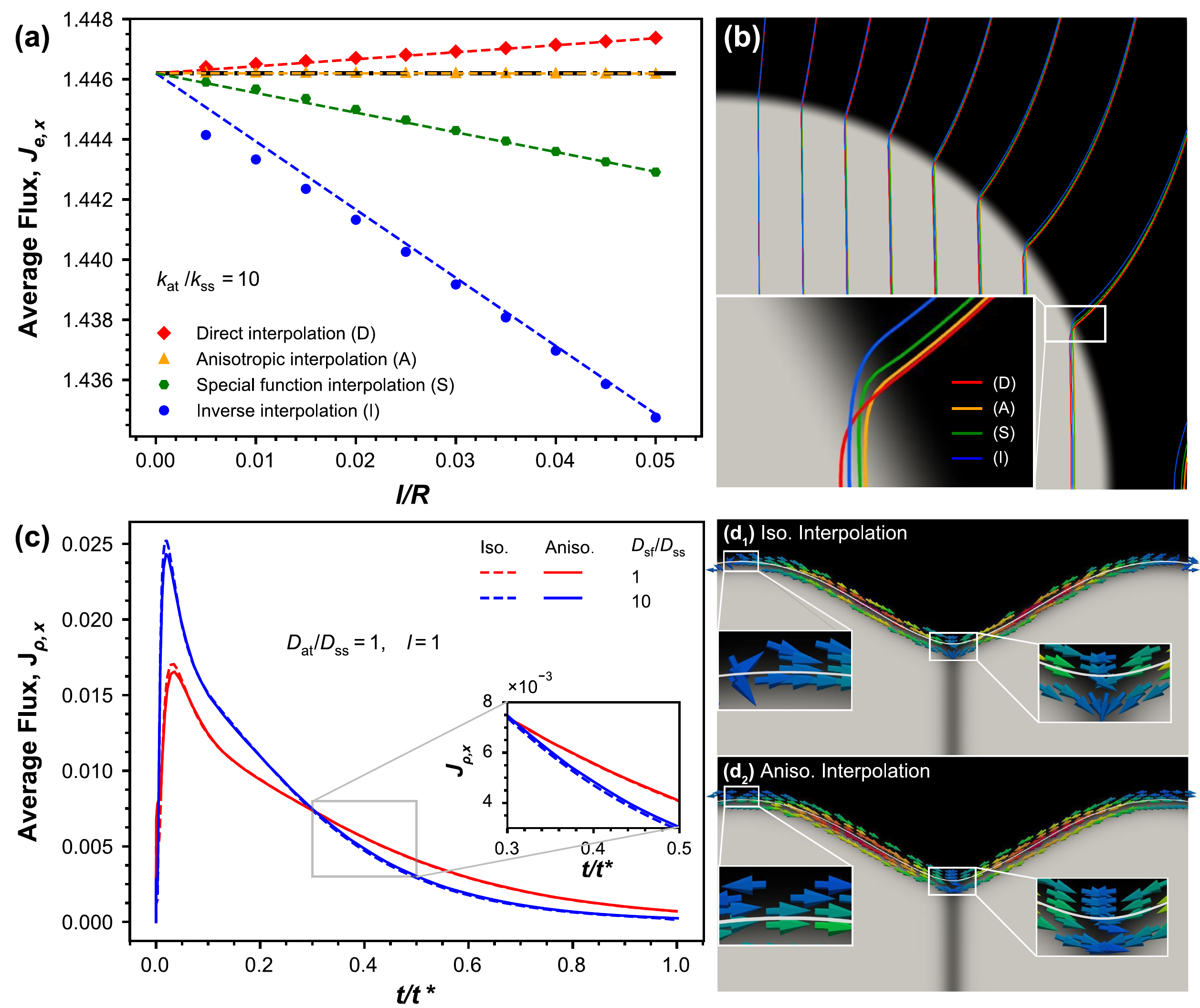}
\caption{\label{fig:interpolation} (a) Comparison of ${J}_{e,x}$ vs $l/R$ for different thermal conductivity interpolations: direct interpolation (red diamonds), inverse interpolation (blue circles), special function interpolation (green hexagons) and anisotropic interpolation (orange triangles) for ${k}_\at/{k}_\ss = 10$ where corresponding color lines are fitted simulation data lines and black dashdot lines represent the plot of a case where no artificial interface effects exist. (b) Temperature isolines across the free surface for different thermal conductivity interpolations at $l/R=0.05$ (c) Time evolution of ${J}_{\rho,x}$ using different diffusivity interpolations: isotropic (dashdot lines), anisotropic (solid lines) for varied values of $D_\text{sf}/D_\ss$. Local zooms around the neck of the grains using ($\mathrm{d_1}$) isotropic interpolation of diffusivity ($\mathrm{d_2}$) anisotropic interpolation of diffusivity. $t^{*} = 14 \times 10^{3}$ unit.} 
\end{figure*}

\subsection{Importance of anisotropic interpolations of the mobility tensor}
Here, we particularly demonstrate the importance of the anisotropic interpolations of the kinetic mobilities.
First, we investigate a steady-state heat transfer case. The numerical validation test proposed by Nicoli et al. \cite{nicoli2011} is used and extended. We consider a square simulation domain defined as $[0,1]$ and $[0,1]$ in $x$ and $y$ direction, respectively and subjected to $\nabla T = -2$ along the x-axis. The domain consists of a stationary disk-shaped solid grain with radius $R$ surrounded by an atmosphere region. A schematic of the simulation setup is supplemented in Fig. S5a.
For ${k}_\at/{k}_\ss = 10$, four cases of thermal conductivity interpolations are examined. We consider the form of interpolation utilized in current phase-field models of non-isothermal sintering \cite{Yang2019, yangscripta2020}. This interpolation form is similar to the formulation given in Eq. \eqref{an10} and is thereafter referred to as the direct interpolation. Also, we consider another form of interpolation given in Ref. \cite{Aalilija2021} to ensure heat flux conservation across the interface. This form of interpolation, thereafter referred to as the inverse interpolation, has its formulation as in Eq. \eqref{an9}. Moreover, a form of interpolation proposed by Almgren \cite{Almgren1999} was used in Refs. \cite{Boussinot2017} and \cite{wang2021} for their non-diagonal phase-field models. The interpolation thereafter called the special function (SF) interpolation is also examined and can be expressed as 
\begin{equation}
\frac{1}{k(\rho)} = \bigg(\frac{1}{ 2{k}_\ss} + \frac{1}{2{k}_\at}\bigg) + p_\text{sf}(\rho)\bigg(\frac{1}{2{k}_\ss} - \frac{1}{2{k}_\at}\bigg), \label{an18}
\end{equation}
with
\begin{equation}
 p_\text{sf}(\rho) = (2\rho-1)[1 + 4a\rho(1-\rho)], \label{an19}
\end{equation}
where $a \approx 0.90$ for ${k}_\at/{k}_\ss =$  10. Here, it is important to note that the formulation in Eq. \eqref{an18} is adopted from Ref. \cite{wang2021} because $\phi$ in Ref. \cite{wang2021} varies from 0 to 1 similar to $\rho$. Lastly, we consider the anisotropic form of thermal conductivity proposed in this work as expressed in Eq. \eqref{an8}. Artificial interface effects are quantified by obtaining the average heat flux, ${J}_{e,x}$ in the domain at $x = 1$. The plot of ${J}_{e,x}$ against normalized interface widths $l/R$ are presented in Fig.~\ref{fig:interpolation}a for different interpolation forms. The black line in Fig.~\ref{fig:interpolation}a indicates a reference case where no artificial interface effect exist, i.e ${J}_{e,x}\mid_{l = 0} \:=\: {J}_{e,x}\mid_{l > 0}$. Fig.~\ref{fig:interpolation}b shows the temperature isolines across the free surface for the different interpolations forms. As shown in Fig.~\ref{fig:interpolation}a, the direct and inverse interpolations show significant deviations from the reference case implying the deficit of these interpolation forms in eliminating interface effects. The SF interpolation also shows considerable deviation from the reference case. This deviation, which might be attributed to the nonmonotonic form of $p_\text{sf}(\rho)$ \cite{ohno2016} reinforces the limitation of the SF interpolation. On the other hand, the results by using the anisotropic interpolation shows very convincing agreement. The outstanding performance of the anisotropic form of interpolation necessitates its consideration for subsequent non-diagonal phase-field modeling. 
Note that for common sintering scenarios, (result is supplemented in Fig. S5b for ${k}_\at/{k}_\ss = 0.2$), the SF interpolation might be utilized for quantitative simulations. The anisotropic form of interpolation, however, finds great importance in other processes where ${k}_\at/{k}_\ss$ is higher such as the case studies in Ref. \cite{yangscripta2022}.

Also, we consider mass transport during grain coalescence of two identical spheres using two mass diffusivity interpolations. A full schematic of the simulation setup is supplemented in Fig. S4b. We make comparisons between the anisotropic interpolation presented in this work, Eq. \eqref{an11} and an isotropic interpolation expressed as \cite{wang2006}:
\begin{equation}
D = p_\ss(\rho){D}_\ss + p_\at(\rho){D}_\at + p_\text{sf}(\rho) D_\text{sf} + p_\text{gb}(\eta_i) D_\text{gb},
\end{equation}
where $p_\ss(\rho)$ and $p_\at(\rho)$ are interpolation functions valued as one only in the solid phase and atmosphere region respectively. Average mass flux ${J}_{\rho,x}$ is obtained across a grain with the plots of ${J}_{\rho,x}$ against normalized time $t$ shown in Fig.~\ref{fig:interpolation}c.

The thin-interface limit analysis showed that eliminating chemical potential jump across the free surface does not require a specific mass diffusivity interpolation. Correspondingly, it has been derived in Refs. \cite{Gugenberger2008, ahmed2016} that the Cahn-Hilliard equation recovers the sharp-interface limit equation of motion for surface diffusion regardless of mass diffusivity form. Therefore, the ${J}_{\rho,x}$ vs $t$ curve is expected to be the same for both interpolations of mass diffusivity, since theoretically no artificial interface effect is related to the diffusivity interpolation.
However, as shown in Fig.~\ref{fig:interpolation}c, there exist surprisingly ${J}_{\rho,x}$ numerical deviations. This can be explained by close comparison of flux details at the free surface region.  Fig.~\ref{fig:interpolation}$\mathrm{d_1}$ and Fig.~\ref{fig:interpolation}$\mathrm{d_2}$ demonstrate the calculated flux by using the isotropic and anisotropic diffusivity form, respectively. We observe that the anisotropic form of diffusivity delivers more reasonable description of the directions of the fluxes. Around the free surface in Fig.~\ref{fig:interpolation}$\mathrm{d_1}$ where we used isotropic diffusivity, there exist non-tangential fluxes at the free surface where only tangential fluxes are expected to contribute to surface diffusion. On the other hand, in Fig.~\ref{fig:interpolation}$\mathrm{d_2}$ where the anisotropic diffusivity form is used, only fluxes that are tangential to the free surface region exist to describe surface diffusion. Accordingly, it is imperative that, while asymptotic analysis confers no restriction on the diffusivity form as regards effecting quantitative simulations in mass diffusion, the anisotropic diffusivity form however makes it possible that the directions of fluxes are effectively described analogous to the sharp-interface description.

\section{\label{sec:conclusions}Conclusions}
In this work, we have developed a variational
quantitative phase-field model for non-isothermal sintering processes following the non-diagonal phase-field approach introduced in Refs. \cite{brener2012, Fang2013}. The model was formulated to eliminate artificial interface effects due to the diffuse-interface description of the free surfaces. Moreover, model formulations are derived in a variational manner guaranteeing thermodynamic consistency. The proposed model differs from conventional non-isothermal sintering models owing to that fact that cross-coupling terms between conserved kinetics (mass and heat transfer) and the non-conserved kinetics (grain growth) are taken into the account. These terms parameterized by functions $M_1$ and $M_2$ can be likened to antitrapping currents in quantitative phase-field modeling. The above-mentioned terms are particularly essential for correct projection of the model to its sharp-interface descriptions. Also, we derive formulations of $M_1$ and $M_2$ in terms of the model parameters using an asymptotic analysis procedure presented in Ref \cite{Boussinot2013}. In addition we showed that anisotropic interpolations of kinetic mobilities are also important to ascertain the elimination of artificial interface effects at the free surface.

Numerical tests were done to highlight the importance of these cross-coupling terms. The results presented showed the emergence of chemical potential jump ($\delta \mu$) and temperature jump ($\delta T$) at the free surface when $M_1 = 0$ and $M_2 = 0$.  $\delta \mu \neq 0$ and $\delta T \neq 0$ negates the sharp-interface sintering description. However, employing $M_1 \neq 0$ and $M_2 \neq 0$ as described in quantitative model eliminates these jumps. The convergence behavior of $\delta \mu$ in respect to interface width ($l$) was presented for model with $M_1 = 0$ and model with $M_1 \neq 0$. For both models, $\delta \mu \to 0$ as $l \to 0$ demonstrating their efficacy at relatively smaller $l$. The major usefulness of quantitative model is seen as $l \gg 0$ where $\delta \mu$ is significantly large for model with $M_1 = 0$ compared to model where $M_1 \neq 0$. Additionally, the difference in transient microstructure and temperature profiles were examined for model with antitrapping currents and model without antitrapping currents. It was seen that the antitrapping currents helps to eliminate extra driving forces brought about by $\delta \mu \neq 0$ and $\delta T \neq 0$ at the free surface.
{Moreover, it was demonstrated that the antitrapping currents only modify the sintering kinetics and have no impact on the thermodynamic conditions.}

Furthermore, we demonstrated numerically how the anisotropic interpolation of kinetic mobilities delivers effective description of diffusion fluxes comparable to sharp-interface description. Therefore, the proposed model can serve as a great tool in studying quantitative simulations of non-isothermal sintering and other related solid-state processes. A major outlook of this work is to further investigate the convergence of interface velocity with respect to interface width obtained using proposed model.

\begin{acknowledgments}
Authors acknowledge the financial support of German Science Foundation (DFG) in the Priority Program 2256 (SPP 2256, project number 441153493) and Collaborative Research center Transregio 270 (CRC-TRR 270, project number 405553726, sub-projects A06). The authors also greatly appreciate their access to the Lichtenberg High-Performance Computer and the technique supports from the HHLR, Technische Universit\"at Darmstadt.
\end{acknowledgments}

\section*{DATA AVAILABILITY}
The authors declare that the data supporting the findings of this study are available within the paper. Source codes of MOOSE-based application NIsoS and related utilities are provided in the online repository \url{bitbucket.org/mfm_tuda/nisos.git}.

\appendix*
\setcounter{figure}{0}   
\setcounter{table}{0} 
\setcounter{equation}{0} 
\renewcommand\thefigure{A\arabic{figure}}
\renewcommand\thetable{A\arabic{table}}
\renewcommand\theequation{A\arabic{equation}}

\section{Deviation of conserved order parameter}

\begin{figure*}
\includegraphics[width=\textwidth]{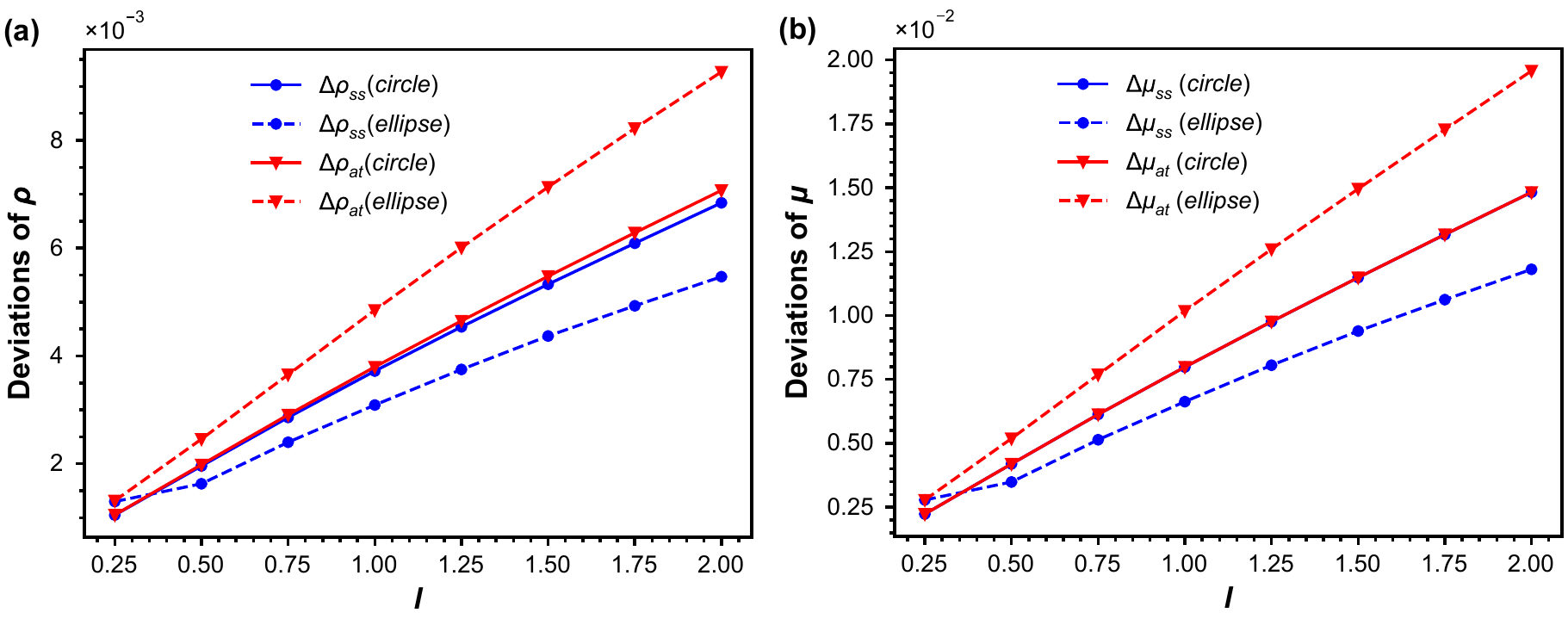}
\caption{\label{fig:deviations} The deviated bulk values of (a) $\rho$ and (b) $\mu$ with respect to the diffuse interface width $l$. The deviated bulk values are read from the numerical results presented in Fig. \ref{fig:ellipse}a.} 
\end{figure*}

{The analyses of Cahn-Hilliard dynamics in Refs. \cite{yue2007} and \cite{dadvand2021} have showed that usage of finite interface width combined with comparable curvature radius induces deviation of the conserved order parameter ($\rho$ in this work) in the bulk regions. It has been demonstrated that the equilibrium bulk values of $\rho$ are contingent on the interface having negligible volume compared to the bulk region, so that only the local free energy finds minimization. Although this condition is viable for planar interfaces, it is not maintained for curved interfaces with concentrated energy. In this sense, the total free energy can be reduced by shrinking the area enclosed by the interface, which subsequently shifts the bulk values of $\rho$ from the equilibrium ones due to the finite volume precept \cite{yue2007}. 
Here, we define the deviated quantities of $\rho$ from its equilibrium values (in this work $\rho_\ss^\eq=1$ and $\rho_\at^\eq=0$) as $\Delta\rho_\ss = \rho_\ss - 1$ and $\Delta\rho_\at = \rho_\at$. Both $\rho_\ss$ and $\rho_\at$ are read from numerical results in Fig.~\ref{fig:ellipse}a with $M_1 = 0$ when the particle is in elliptical and circular shapes. The tendencies of $\Delta \rho_\ss$ and $\Delta \rho_\at$ vs. $l$ are respectively shown in Fig.~\ref{fig:deviations}a. Similar to results obtained in Ref. \cite{yue2007}, $\Delta \rho_\ss$ and $\Delta \rho_\at$ increase with increasing $l$. When in the elliptical shape (implying a non-equilibrium condition), $\Delta \rho_\at > \Delta \rho_\ss$ holds for almost every selected $l$, while $\Delta \rho_\ss \approx \Delta \rho_\at$ when in the circular shape (implying an equilibrium condition). These differences can be attributed to the curvature dependency of the analytical profile of $\rho$ \cite{dadvand2021}. 
It should be noted that $\Delta \rho_\ss$ and $\Delta \rho_\at$ exist even for symmetric mobilities with sufficiently large $l$, which is distinctive to the known interface effects (like trap effects) that are incited by asymmetric kinetic mobilities. Moreover, the comparison of analytical values of $\Delta\rho_\ss$ and $\Delta\rho_\at$ and numerical values should be examined in further studies.}

{As one of the significant outcomes, deviated bulk values of $\rho$ incite deviated chemical potential $\mu$ from its equilibrium ones in the bulk regions, which may result in the unexpected chemical potential drop as an extra driving force across the free surface. To examine this point, we define the deviated quantities of $\mu$ in a similar fashion as $\Delta \rho_\ss$ and $\Delta \rho_\at$, i.e., $\Delta \mu_\ss = \mu_\ss(\rho_\ss)-\mu^\eq = \mu_\ss(\rho_\ss)$ and $\Delta \mu_\at = \mu_\at(\rho_\at)-\mu^\eq_\at = \mu_\at(\rho_\at)$, noting that $\mu^\eq_\ss(\rho^\eq_\ss=1)=\mu^\eq_\at(\rho^\eq_\at=0)=0$. In Fig.~\ref{fig:deviations}b, we present a similar tendency of $\Delta \mu_\ss$ and $\Delta \mu_\at$ vs. $l$ when the particle is in the elliptical shape, where both $\Delta \mu_\ss$ and $\Delta \mu_\at$ grow along with increasing $l$ and $\Delta \mu_\ss > \Delta \mu_\at$ is depicted for every selected $l$, implying the existing chemical potential drop $\Delta \mu = \Delta \mu_\at-\Delta \mu_\ss >0$ across the free surface at the semi-major axis, as shown in Fig.~\ref{fig:ellipse}a. Notably, when the particle is in the circular shape, $\Delta \mu_\ss = \Delta \mu_\at$ is formed without the dependency of $l$, indicating no chemical potential drop across the free surface, i.e., $\Delta \mu =\Delta \mu_\at-\Delta \mu_\ss=0$. This also demonstrates that the existing deviation in $\mu$ incited by $\Delta \rho_\ss$ and $\Delta \rho_\at$ does not affect the supposing equilibrium condition, as the particle stops morphing in the circular shape.}

\begin{figure*}
\includegraphics[width=\textwidth]{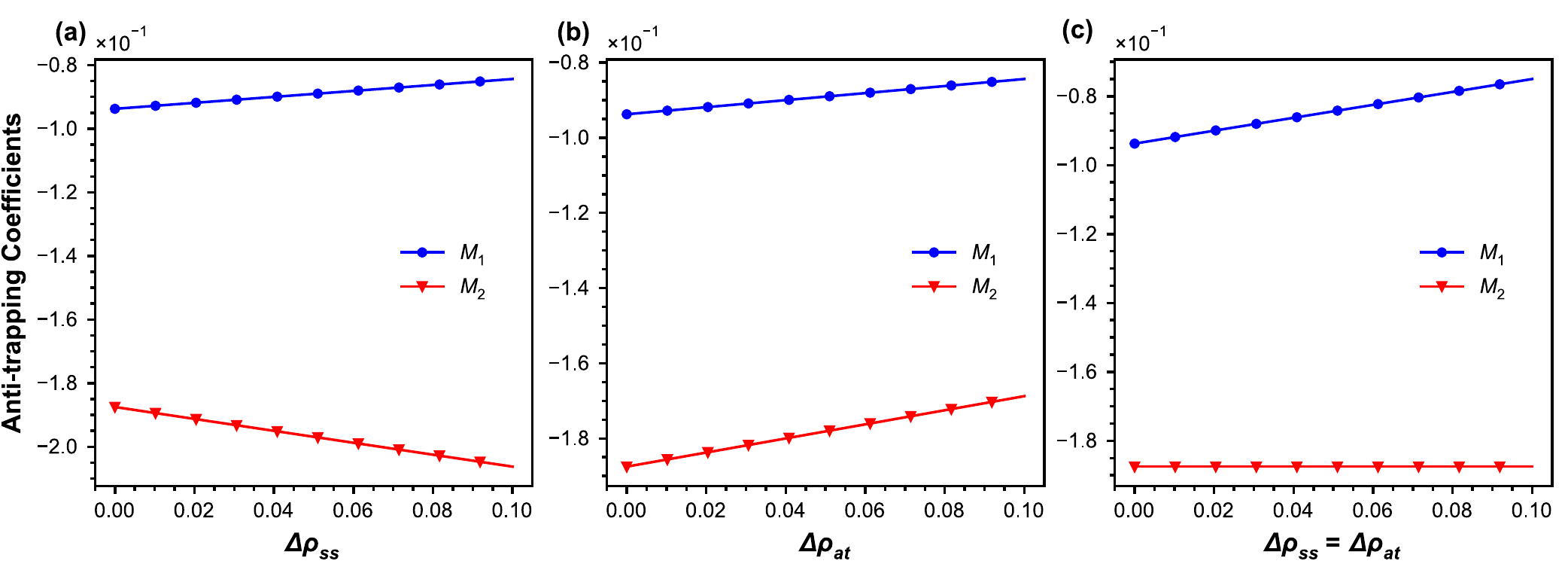}
\caption{\label{fig:ATdeviations} The antitrapping coefficients $M_1$ and $M_2$ with respect to (a) varying $\Delta \rho_\ss$ when $\Delta \rho_\at =0$, (b) varying $\Delta \rho_\at$ when $\Delta \rho_\ss = 0$, and (c) varying simutaneously $\Delta \rho_\ss$ and $\Delta \rho_\at$ while holding $\Delta \rho_\ss=\Delta \rho_\at$.} 
\end{figure*}

{Additionally, since the antitrapping coefficients $M_1$ and $M_2$ are dependent on the bulk values $\rho_\ss$ and $\rho_\at$, we examine the variations of $M_1$ and $M_2$ with $\Delta \rho_\ss$ and $\Delta \rho_\at$ up to 0.1, as seen in Fig.~\ref{fig:ATdeviations}. In Fig.~\ref{fig:ATdeviations}a, we take $\Delta \rho_\at = 0$ and examine the variations of $M_1$ and $M_2$ with $\Delta \rho_\ss$. Similarly, we take $\Delta \rho_\ss = 0$ and examine the variations of $M_1$ and $M_2$ with $\Delta \rho_\at$ in Fig.~\ref{fig:ATdeviations}b. Then, we present the variations of $M_1$ and $M_2$ with $\Delta \rho_\ss = \Delta \rho_\at$ in Fig.~\ref{fig:ATdeviations}c. It demonstrates that $M_1$ presents a linear tendency vs. increasing deviations of all cases. $M_2$,  however, decreases along with growing $\Delta \rho_\ss$ but increases with growing $\Delta \rho_\at$. For $\Delta \rho_\ss = \Delta \rho_\at$, $M_2$ stays constant. This can be explained via Eq.~(\ref{an15}) where $M_2$ is proportional to $(\rho_\ss - \rho_\at)$, which 
is reduced to one when $\Delta \rho_\ss = \Delta \rho_\at$ as $(\rho_\ss - \rho_\at)=[(\Delta\rho_\ss+1)-\Delta\rho_\at]=1$.}

\providecommand{\noopsort}[1]{}\providecommand{\singleletter}[1]{#1}%

\end{document}